\documentclass[%
 reprint,
superscriptaddress,
 amsmath,amssymb,
 aps,
pra,
showkeys
]{revtex4-2}

\usepackage{graphicx}
\usepackage{subcaption}
\usepackage{dcolumn}
\usepackage{bm}
\usepackage{dsfont}
\usepackage{hyperref}
\usepackage{braket}
\usepackage{mathtools}
\usepackage[mathlines]{lineno}
\usepackage{amsmath,amssymb,nccmath}
\usepackage{xcolor}
\usepackage{mdframed}

\newcommand{\Hamil}{\mathcal{H}}
\newcommand{\Y}{Y}
\newcommand{\mean}[1]{\left\langle#1\right\rangle}

\newcommand{\doubleketbra}[2]{\left|#1\middle\rangle\!\middle\rangle\!\middle\langle\!\middle\langle#2\right|}
\newcommand{\doubleket}[1]{\left|#1\middle\rangle\!\right\rangle}
\renewcommand{\vec}[1]{\bm{#1}}

\newmdenv[
  topline=false,
  bottomline=false,
  rightline=false,
  skipabove=\topsep,
  skipbelow=\topsep
]{leftrule}

\begin{document}

\title{\bf Currents in non-equilibrium steady states of open inhomogeneous $XX$-spin chains}

\author{Pierre-Antoine Bernard}
 \email{bernardpierreantoine@outlook.com}
 \affiliation{Centre de Recherches Math\'ematiques, Universit\'e de Montr\'eal, P.O. Box 6128, Centre-ville Station, Montr\'eal (Qu\'ebec), H3C 3J7, Canada.}
 
\author{Ismaël Bussière}
 \email{ismael.olivier.pacifique.bussiere@umontreal.ca}
 \affiliation{Centre de Recherches Math\'ematiques, Universit\'e de Montr\'eal, P.O. Box 6128, Centre-ville Station, Montr\'eal (Qu\'ebec), H3C 3J7, Canada.}

\author{Roberto Floreanini}
\affiliation{Istituto Nazionale di Fisica Nucleare (INFN), Sezione di Trieste, I-34151 Trieste, Italy}

\author{Luc Vinet}
 \affiliation{Centre de Recherches Math\'ematiques, Universit\'e de Montr\'eal, P.O. Box 6128, Centre-ville Station, Montr\'eal (Qu\'ebec), H3C 3J7, Canada.}
 \affiliation{IVADO, 6666 Rue Saint-Urbain, Montr\'eal (Qu\'ebec), H2S 3H1, Canada}

\date{August 9, 2024}

\begin{abstract}
   
We investigate spin and heat currents in the non-equilibrium steady state of inhomogeneous $XX$-spin chains, which act as interfaces between two bosonic heat baths. Using special functions that diagonalize the single-particle Hamiltonian, we derive closed-form expressions for these currents. For small temperature differences between the baths, we show that inhomogeneities breaking the mirror symmetry of the chain significantly reduce both heat and spin conductivities. Connections with perfect state transfer are discussed. 
\end{abstract}

\maketitle


\section{Introduction}

Spin chains offer a convenient testbed to study how microscopic physical features of quantum systems can impact their macroscopic properties. Due to their inherent simplicity and the diverse array of phenomena they showcase, these models have become a well established tool for studying thermalization \cite{steinigeweg2013eigenstate,deutsch1991quantum,srednicki1994chaos}, entanglement in quantum many-body systems \cite{calabrese2004entanglement, eisert2010colloquium,calabrese2005evolution} and non-equilibrium steady states in open settings \cite{prosen2009matrix, prosen2008third, prosen2011open,de2018hydrodynamic,vznidarivc2016diffusive}.

Non-equilibrium steady states emerge in systems acting as interfaces between reservoirs or heat baths at different temperatures. These states exhibit non-zero expectations of heat and spin currents, making them very useful for modeling the contribution of magnetic excitations to the thermal conductivity of specific materials (e.g. see \cite{sologubenko2007thermal, pan2022unambiguous,hlubek2010ballistic} for experimental realizations).  As a result, there is much interest in examining how currents in these states scale with the chain length and the temperatures of the driving baths. 

Various factors significantly influence the scaling, including anisotropy \cite{karrasch2015spin}, inhomogeneous (random) couplings \cite{zhang2008ballistic, laflorencie2004crossover}, magnetic fields \cite{heidrich2005thermal}, the presence of dephasing \cite{mendoza2013heat}, and combinations thereof \cite{lange2019driving, lenarvcivc2015exact, schulz2018energy, vznidarivc2013transport}. Additionally, some spin chains exhibit non-zero rectification, demonstrating asymmetric conductivity \cite{landi2014flux}.

Several approaches have been employed to investigate the effects of these microscopic features. Notable methods include linear response theory \cite{klumper2002thermal, heidrich2005thermal}, the study of steady states within the light cone induced by an inhomogeneous quench \cite{langer2009real,biella2019ballistic}, and the construction of solutions to Lindblad equations (e.g. \cite{prosen2011open,prosen2009matrix}). Using the latter, recent works have studied  non-equilibrium steady states in homogeneous $XX$-spin chains of size $N$ sitting in between bosonic heat baths \cite{benatti2021exact,benatti2022stationary}. The authors derived the Lindblad equation governing the reduced dynamics and obtained an explicit formula for the density matrix associated with the non-equilibrium steady state. This led to the identification of source and sink terms that characterize the spin and energy currents within the chain.

In this paper, we adopt a similar approach to investigate the conductivity of $XX$-spin chains with inhomogeneous couplings and magnetic fields. While some results exist for heat and spin currents in systems with random, localized, and alternating inhomogeneities, a comprehensive understanding of the effects of general inhomogeneities remains elusive.  
A notable unanswered question is whether inhomogeneous chains designed for perfect quantum state transfer \cite{christandl2004perfect, howToConstruct} exhibit a particular tendency to support spin currents when their ends are coupled to the environment. Originally introduced in quantum information as potential \textit{quantum wires} moving qubits with minimal intervention through their intrinsic dynamics, it is unclear how this property translates to spin conductivity in an open setting.

To address this issue, we focus on the specific case of gapped $XX$-spin chains coupled to bosonic baths, as detailed in \cite{benatti2021exact, benatti2022stationary}. This model allows for analytical results based on the single-particle dynamics of the system. In particular, we derive formulas that relate spin and heat currents to the special functions used to diagonalize the one-particle Hamiltonian. Our analysis yields several significant observations in different regimes, including:
\begin{enumerate}
    \item[(i)] \textit{Small temperature gap, low temperature}: Heat and spin flow vanish exponentially as the temperature approaches $0$ due to the gap. The scaling of the currents with the system size $N$ is influenced by the wavefunction of the low energy single excitation states.

    \item[(ii)] \textit{Small temperature gap, high temperature}: Any spin chain that is symmetric with respect to its center exhibits ballistic transport in this regime. However, this property is fragile; even a small perturbation of the magnetic field can cause a transition to subdiffusive transport.
    
    \item[(iii)] \textit{High temperature gap}: The inhomogeneities in the spin chains have minimal impact on the currents in this regime. The heat current is proportional to the energy of a spin-up located at the site adjacent to the bath at the lower temperature, which serves as the exiting bath.

\end{enumerate}
The paper is structured as follows. In Section \ref{sec:2a}, we recall the definition and diagonalization of the Hamiltonian for inhomogeneous spin chains of type \( XX \). Two examples are provided in Section \ref{sec:2b}. Thermal baths interacting with the extremities of the chain are introduced in Section \ref{sec:2c}, along with the derivation of the associated Lindblad equation. An explicit expression for the density matrix of the non-equilibrium steady state is derived in \ref{sec:3}. The associated spin and thermal currents are studied in Section \ref{sec:4}.

\section{Lindblad equation}\label{sec:2}

\subsection{Closed model}\label{sec:2a}
We consider inhomogeneous spin chains of type $XX$ defined by the following Hamiltonian,
\begin{equation}\label{eq:H1}
    \Hamil=\frac{1}{2}\sum_{\ell=0}^{N-1} J_\ell\left(\sigma^x_\ell\sigma^x_{\ell+1}+\sigma^y_\ell\sigma^y_{\ell+1}\right)+\frac{1}{2}\sum_{\ell=0}^N (B_{\ell}+\Delta)\left(1+\sigma^z_\ell\right),
\end{equation}
where $J_n$ and $B_n$ are nearest-neighbor couplings and local magnetic field strengths, respectively and $\sigma_\ell^x,\sigma_\ell^y,\sigma_\ell^z$ are Pauli matrices acting on site $\ell$. The parameter $\Delta$ is a constant representing the strength of a constant background magnetic field. 
By means of a Jordan-Wigner transformation,
\begin{equation}
    a^\dagger_n=e^{-i\pi\sum_{\ell=0}^{n-1}\sigma^+_\ell\sigma^-_\ell}\sigma^+_n,
\end{equation}
where $\sigma_\ell^\pm=\frac{1}{2}\left(\sigma^x_\ell\pm i\sigma_\ell^y\right)$, the Hamiltonian \eqref{eq:H1} can be recast as a model of free fermions hopping on a chain,
\begin{equation}
    \Hamil=\sum_{n=0}^{N-1}J_n\left(a^\dagger_n a_{n+1}+a^\dagger_{n+1}a_n\right)+\sum_{n=0}^N(B_n+\Delta)a^\dagger_na_n,
\end{equation}
where $a^\dagger_n$ and $a_n$ are fermionic creation and annihilation operators satisfying $\left\{a^\dagger_{n},a_{m}\right\}=\delta_{nm}$ and $\left\{a_n,a_m\right\}=0$. The diagonalization of this Hamiltonian requires considering its restriction to the subspace of single excitations. Let us denote $\doubleket{\vec{0}}$ the state where all spins are down, which is annihilated by all operators $a_n$,
\begin{equation}
    a_n\doubleket{\vec 0}=0,\ \forall n \in \{0,1,\dots, N\}
\end{equation}
The action of the Hamiltonian $\Hamil$ on states $\ket{n} = a_n^\dagger \doubleket{\vec 0}$ with a single spin up at position $n$ is
\begin{equation}
    \Hamil\ket{n}=J_n\ket{n+1}+B_n\ket{n}+J_{n-1}\ket{n-1}
\end{equation}
which corresponds to a tridiagonal matrix $\mathrm H + \Delta$, with
\begin{equation}\label{eq:hamiltonian-matrix}
    \mathrm H=\begin{pmatrix}
        B_0   &J_0   &0     &\dots  &0      \\
        J_0   &B_1   &J_1   &\ddots &\vdots \\
        0     &J_1   &B_2   &\ddots &0      \\
        \vdots&\ddots&\ddots&\ddots &J_{N-1}\\
        0     &\dots &0     &J_{N-1}&B_N
    \end{pmatrix}
\end{equation}
The spectral problem for $\mathrm H$ is solved by introducing wavefunctions $\phi_n(x_k)$ that satisfy the three term recurrence relation
\begin{equation}
    x_k\phi_n(x_k)=J_n\phi_{n+1}(x_k)+B_n\phi_n(x_k)+J_{n-1}\phi_{n-1}(x_k)\ ,
\end{equation}
and the orthonormality conditions,
\begin{equation}
\sum_{n=0}^N\phi_n(x_k)\phi_n(x_{k'})=\delta_{kk'},\quad \sum_{k=0}^N\phi_n(x_k)\phi_m(x_{k})=\delta_{nm}\ .
\end{equation}
These relations can be understood in terms of an orthogonal transition matrix $U$ with $U_{nk}=\phi_n(x_k)$ as
\begin{equation}
    \text{diag}(x_k+\Delta)=U^THU
\end{equation}
and
\begin{equation}
    UU^T=1,\qquad U^TU=1
\end{equation}
It clearly follows that the $\phi(x_k)$ are the wavefunctions of single-excitation stationary states with energy $x_k + \Delta$,
\begin{equation}
    (\mathrm{H} + \Delta) \sum_n \phi_n(x_k) \ket{n} = (x_k+ \Delta) \sum_n \phi_n(x_k) \ket{n}\ .
\end{equation}
These wavefunctions can always be expressed in terms of discrete orthogonal polynomial families as will be discussed in section \ref{sec:2b}.
Furthermore, they allow the Hamiltonian 
$\Hamil$ to be rewritten in a diagonal form as
\begin{equation}
    \Hamil=\sum_{k=0}^N(x_k+\Delta)b_k^\dagger b_k
\end{equation}
where $b_k=\sum_{n}\phi_n(x_k)a_n$ and $b_k^\dagger=\sum_{n}\phi_n(x_k)a_n^\dagger$ are new pairs of fermionic creation and annihilation operators. A complete eigenbasis is obtained by the application of creation operators $b^\dagger_k$ on the state $\doubleket{\vec 0}$, with eigenvectors labeled by a binary vector $\vec n$ representing the empty and occupied modes,
\begin{equation}
    \doubleket{\vec n}=\prod_{k=0}^{N}(b_k^\dagger)^{\vec n_k}\doubleket{\vec 0}
\end{equation}
These eigenstates have energies $E_{\vec n}=\sum_{k=0}^N\vec n_k\left(x_k+\Delta\right)$. In the following, we restrict ourselves to the case where the background magnetic field is large enough so that all modes have positive energy $x_k + \Delta > 0$, making $\doubleket{\vec 0}$ the ground state.

The choice of different inhomogeneous coefficients $J_\ell$ and $B_\ell$ affects only the energies $x_k + \Delta$ and wavefunctions $\phi_n(x_k)$ of the single-excitation stationary states. This is illustrated in the following subsections with two examples.
\subsection{Two examples}\label{sec:2b}
Since the wavefunctions $\phi_n(x_k)$ are solutions to a three-term recurrence relation, Favard's theorem implies that they can be expressed in terms of orthogonal polynomials. Consequently, the recurrence coefficients of known families of orthogonal polynomials provide examples of inhomogeneous spin chains that can be exactly diagonalized. As examples, we consider the homogeneous and Krawtchouk chains. The former is characterized by
\begin{equation}
    J_n=1,\quad  B_n=0 \ .
\end{equation}
The associated single excitation energy spectrum $x_k$ and wavefunctions $\phi_n(x_k)$ are well-known \cite{crampe2021entanglement} to be given by
\begin{align}
     x_k&=\cos\left(\frac{(k+1)\pi}{N+2}\right) \nonumber
\end{align}
and
\begin{align}
        \phi_n(x_k)&=\sqrt{\frac{2}{N+2}}\sin\left(\frac{(k+1)\pi}{N+2}\right)U_n(x_k) \nonumber\\
        &=\sqrt{\frac{2}{N+2}}\sin\left(\frac{(n+1)(k+1)\pi}{N+2}\right),
\end{align}
where $U_n(x)$ are the Chebyshev polynomials of the second kind.
\\
The Krawtchouk spin chain \cite{christandl2004perfect, howToConstruct} is characterized by nearest-neighbor couplings that peak at the center of the chain and a linearly increasing magnetic field,
\begin{gather}
    \begin{gathered}
        J_n=\sqrt{p(1-p)}\sqrt{(n+1)(N-n)}, \\
        B_n=p(N-n)+(1-p)n \ ,
    \end{gathered}
\end{gather}
with $p$ as a free parameter in the interval $]0,1[$. The associated single excitation energy spectrum $x_k$ and wavefunctions $\phi_n(x_k)$ are
\begin{gather}
    \begin{gathered}
        x_k=k 
    \end{gathered}
\end{gather}
and
\begin{gather}
    \begin{gathered}
        \phi_n(x_k)=\sqrt{\begin{pmatrix}
            N\\ k
        \end{pmatrix}\begin{pmatrix}
            N\\ n
        \end{pmatrix}p^{k+n}(1-p)^{N-k-n}}K_n(k;p,N)
    \end{gathered}
\end{gather}
where $K_n(x;p,N)$ are the Krawtchouk polynomials \cite{koekoek}. \\

Note that both the homogeneous chain and the Krawtchouk chain with $p=1/2$ exhibit mirror symmetry, meaning they have coefficients such that  $J_n=J_{N-n-1}$ and $B_n=B_{N-n}$. This mirror symmetry is a necessary (but not sufficient) condition for the chain to achieve perfect state transfer \cite{kay2010perfect, howToConstruct}. Perfect state transfer occurs when a single spin-up localized at one end of the chain is transferred to the opposite end at a specified time $T$ with probability $1$, i.e.
\begin{equation}
    |\bra{0} e^{-i T H} \ket{N}| = 1.
\end{equation}
This phenomenon does not occur in the homogeneous chain but does occur at time 
$T = \pi$ in the Krawtchouk chain with 
$p = 1/2$ \cite{albanese2004mirror}.

\subsection{Open model}\label{sec:2c}
We recall the derivation from \cite{benatti2021exact} for the Lindblad equation associated with an $XX$ spin chain which interacts with two auxiliary bosonic thermal bath described by the following Hamiltonians:

\begin{equation}
    \Hamil_\alpha=\int_0^\infty\!\mathrm{d}\nu\ \nu \mathfrak{b}_\alpha^\dagger(\nu)\mathfrak{b}_\alpha(\nu),\qquad \alpha\in\{0,N\}
\end{equation}
where $\mathfrak{b}_\alpha$ and $\mathfrak{b}^\dagger_\alpha$ satisfy the canonical bosonic commutation relations
\begin{equation}
    \left[\mathfrak{b}_\alpha^\dagger(\nu),\mathfrak{b}_\beta(\nu')\right]=\delta_{\alpha\beta}\delta(\nu-\nu').
\end{equation}
The coupling between the baths and the chain is mediated by the following interaction Hamiltonian:
\begin{equation}\label{eq:interaction-hamiltonian}
    \Hamil_I=\sum_{\alpha\in\{0,N\}}\left(\sigma^+_\alpha\mathfrak{B}_\alpha+\sigma^-_\alpha\mathfrak{B}^\dagger_\alpha\right)
\end{equation}
where 
\begin{equation}
    \mathfrak{B}_\alpha=\int_0^\infty\mathrm d\nu\ h_\alpha(\nu)\mathfrak{b}_\alpha(\nu)
\end{equation}
with $h_\alpha(\nu)$, real suitable smearing functions. This interaction Hamiltonian is appropriate only when all energies of the closed spin chain are positive. A more general interaction Hamiltonian as been considered for the homogeneous chain in \cite{benatti2022stationary}. \\
For the case we are studying, the complete Hamiltonian is
\begin{equation}
    \Hamil_\mathrm{tot}=\Hamil+\lambda \Hamil_I+\Hamil_0 + \Hamil_N
\end{equation}
where $\lambda\ll 0$ is a dimensionless coupling parameter. \\
We consider the two baths to be at thermal equilibirum at temperature $T_\alpha=1/\beta_\alpha$ so that we can write their density matrices as
\begin{equation}
    \rho_\alpha=\frac{e^{-\beta_\alpha \Hamil_\alpha}}{\mathrm{Tr}\left(e^{-\beta_\alpha \Hamil_\alpha}\right)}
\end{equation}
Assuming that there is weak coupling between the chain and the baths, the baths remain in their thermal state which means the density matrix of the total system can be written as
\begin{equation}
    \rho_\mathrm{tot}(t)=\frac{e^{-\beta_0 \Hamil_0}}{\mathrm{Tr}\left(e^{-\beta_0 \Hamil_0}\right)}\otimes\rho(t)\otimes\frac{e^{-\beta_N \Hamil_N}}{\mathrm{Tr}\left(e^{-\beta_N \Hamil_N}\right)}
\end{equation}
where $\rho(t)$ is the reduced density matrix that carries the dynamics of the chain. From the Von Neumann equation
\begin{equation}
    \frac{\mathrm d}{\mathrm dt}\rho_\mathrm{tot}(t)=-i\left[\Hamil_\mathrm{tot},\rho_\mathrm{tot}(t)\right],
\end{equation}
assuming weak coupling and perfoming the secular approximation, one can derive the Lindblad master equation (as is done in the Appendix) which governs the evolution of the reduced density $\rho(t)$:
\begin{equation}\label{eq:Lindblad1}
    \frac{\mathrm d}{\mathrm dt}\rho(t)=-i\left[\Hamil+\lambda^2H_{LS},\rho(t)\right]+\mathds{D}[\rho(t)]=\mathds{L}[\rho]
\end{equation}
where $\Hamil_{LS}$, the Lamb-shift Hamiltonian corrects the energy levels and $\mathds{D}[\rho(t)]$ is the dissipative contribution to the otherwise unitary evolution of $\rho(t)$. For the open spin chain, the Lamb-shift Hamiltonian is
\begin{equation}\label{eq:clean-HLS}
    \Hamil_{LS}=\sum_{k=0}^N\left(s_kb_k^\dagger b_k+\tilde{s}_kb_kb_k^\dagger\right)
\end{equation}
and the dissipative contribution is 
\begin{align}
    \mathds{D}[\rho]=\lambda^2\sum_{k=0}^N&\left( d_k\left(b_k^\dagger\rho(t)b_k-\frac{1}{2} \left\{\rho(t),b_xb_x^\dagger\right\}\right)\right. \nonumber\\
    \label{eq:clean-diss}
    &+\left.\tilde{d}_k\left(b_k\rho(t)b_k^\dagger-\frac{1}{2} \left\{\rho(t),b_k^\dagger b_k\right\}\right)\right)
\end{align}
where
\begin{gather}\label{eq:def-dk}
    \begin{gathered}
        d_k=\sum_{\alpha\in\{0,N\}}2\pi \phi_\alpha(x_k)^2h_\alpha(x_k + \Delta)^2(n_\alpha(x_k + \Delta)+1) \\
        \tilde{d}_k=\sum_{\alpha\in\{0,N\}}2\pi \phi_\alpha(x_k)^2h_\alpha(x_k + \Delta)^2n_\alpha(x_k + \Delta)
    \end{gathered}
\end{gather}
\begin{gather}
    \begin{gathered}
        s_k=\sum_{\alpha\in\{0,N\}} \phi_\alpha(x_k)^2\mathrm{P.V.}\!\int_0^\infty\!\mathrm{d}\nu\ \frac{h_\alpha^2(\nu)(n_\alpha(\nu)+1)}{x_k + \Delta-\nu} \\
        \tilde{s}_k=\sum_{\alpha\in\{0,N\}} \phi_\alpha(x_k)^2\mathrm{P.V.}\!\int_0^\infty\!\mathrm{d}\nu\ \frac{h_\alpha^2(\nu)n_\alpha(\nu)}{\nu-x_k - \Delta}\ .
    \end{gathered}
\end{gather}
with $n_\alpha(\omega)=\frac{1}{e^{\beta_\alpha \omega}-1}$ .

\section{Non-equilibrium steady state}\label{sec:3}
Due to the dissipative term in the Lindblad equation, any initial state of the chain will converge after sufficient time to a non-equilibrium steady state, whose density matrix $\rho_\infty$ satisfies
\begin{equation}
    -i\left[\Hamil+\lambda^2\Hamil_{LS},\rho_\infty\right]+\mathds{D}[\rho_\infty] = 0.
\end{equation}
The density matrix $\rho_\infty$ was explicitly derived for a homogeneous chain in \cite{benatti2021exact}. We recall this approach here, which also applies to inhomogeneous spin chains. Consider the ansatz given by the matrix $\Y$, which is a sum of eigenstate projectors:
\begin{equation}
    \Y\coloneqq\sum_{\vec n}y_{\vec n}\doubleketbra{\vec n}{\vec n}
\end{equation}
The commutator $\left[\Y,\Hamil+\Hamil_{LS}\right]$ from the Lindbladian acting on $\Y$ is zero because $\Hamil+\Hamil_{LS}$ acts diagonally on these projectors. The action of the dissipator on the eigenstate projectors is
\begin{align}
    \mathds{D}[\doubleketbra{\vec n}{\vec n}]=\lambda^2\sum_{k=0}^N&\left(n_kd_k-(1-n_k)\tilde d_k\right)\times \nonumber \\
    &\left(\doubleketbra{\vec n_{0_k}}{\vec n_{0_k}}-\doubleketbra{\vec n_{1_k}}{\vec n_{1_k}}\right)
\end{align}
where $\vec n_{i_k}$ means the binary vector $\vec n$ with its $k$-th component replaced by $i$.
It follows that
\begin{align}
    \mathds{D}[\Y]=\lambda^2\sum_{k=0}^N\sum_{\hat{\vec{n}}_k}&\left(d_ky_{\vec{n}_{1_k}}-\tilde{d}_ky_{\vec{n}_{0_k}}\right)\times \nonumber \\ \label{eq:diss-of-diag}
    &\left(\doubleketbra{\vec n_{0_k}}{\vec n_{0_k}}-\doubleketbra{\vec n_{1_k}}{\vec n_{1_k}}\right)
\end{align}
where the sum over $\hat{\vec{n}}_k$ means summing over all binary vectors $\vec n$ with the $k$-th component fixed. Looking at the coefficient, one can see that by choosing
\begin{equation}
    y_{\vec n}=\prod_{k=0}^N\left( n_k\tilde d_k+(1-n_k)d_k\right)
\end{equation}
the coefficient in \eqref{eq:diss-of-diag} will vanish. The following density matrix then describes the non-equilibrium steady state of the spin chain
\begin{equation*}
    \rho_\infty=\sum_{\vec n}\lambda_{\vec n}\doubleketbra{\vec n}{\vec n}
\end{equation*}
where
\begin{equation}
    \lambda_{\vec n}=\frac{y_{\vec n}}{\prod_{k=0}^N(d_k+\tilde{d}_k)}
\end{equation}
so that $\sum_{\vec n}\lambda_{\vec n}=1$. \\

It can be shown that for equal bath temperatures and identical smearing functions for the two baths, this is the same as the thermal equilibrium state \cite{benatti2021exact},
\begin{equation*}
    \rho_\infty = \frac{e^{-\beta \Hamil}}{\mathrm{Tr}\!\left(e^{-\beta \Hamil}\right)}\ .
\end{equation*}

\section{Heat and spin currents}\label{sec:4}

We now characterize the spin and heat currents in the non-equilibrium steady state. 

Given a solution $\rho(t)$ of the Lindblad master equation \eqref{eq:Lindblad1}, the associated spin flow in the chain is defined in terms of the time derivative of the total spin operator, i.e.
\begin{equation}\label{eq:spin-current-1}
    \mathfrak Q= \frac{\mathrm d}{\mathrm dt}\sum_n\mean{\sigma^z_n}=\sum_n\frac{\mathrm d}{\mathrm dt}\mathrm{Tr}\left(\sigma^z_n\rho(t)\right)=\sum_n\mathrm{Tr}\left(\sigma^z_n\mathds L[\rho]\right).
\end{equation}
Using the cyclity of the trace, this expression can be rewritten in terms of an \textit{adjoint Lindbladian} $\tilde{\mathds{L}}$ as
\begin{equation}
    \mathfrak Q= \sum_n \mathrm{Tr}\left(\tilde{\mathds L}[\sigma^z_n]\rho\right)  \\
\end{equation}
where
\begin{equation}\label{eq:adjoint-lindblad}
    \tilde{\mathds{L}}[\mathcal{O}]=i[\Hamil+\lambda^2\Hamil_{LS},\mathcal{O}]+\tilde{\mathds{D}}_0[\mathcal{O}] + \tilde{\mathds{D}}_N[\mathcal{O}].
\end{equation}
The operators $\tilde{\mathds{D}}_0$ and  $\tilde{\mathds{D}}_N$ are the \textit{adjoint dissipators}. They respectively capture the dissipative effects due to the baths interacting at site $0$ and $N$, and have the following action on an operator $\mathcal{O}$,
\begin{align}
     \tilde{\mathds D}_\alpha[\mathcal{O}]=\sum_{k}\phi_\alpha(x_k)&\!\left(C_{\alpha,k}\left(b_k^\dagger\mathcal{O} b_k-\frac{1}{2}\left\{b_k^\dagger b_k,\mathcal{O}\right\}\right)\right. \nonumber \\
     \label{eq:dissipRL}
     &\ \left.+\tilde{C}_{\alpha,k}\left(b_k\mathcal{O} b_k^\dagger-\frac{1}{2}\left\{b_k b_k^\dagger,\mathcal{O}\right\}\right)\right),
\end{align}
with coefficients $C_{\alpha,k}$ and $\tilde{C}_{\alpha,k}$ given by
\begin{gather}
    \begin{gathered}
    C_{\alpha,k}=2\pi h_\alpha(x_k + \Delta)^2(n_\alpha(x_k + \Delta)+1) ,\\
    \tilde{C}_{\alpha,k}=2\pi h_\alpha(x_k + \Delta)^2n_\alpha(x_k + \Delta)    
    \end{gathered}\ .
\end{gather}
For currents in the non-equilibrium steady state, the contribution from the commutator vanishes, allowing us to discard it in \eqref{eq:adjoint-lindblad}. The adjoint dissipators separate the contributions from each bath to the spin flow, so that $\mathfrak{Q}$ decomposes into components coming from the left and right bath,
\begin{equation}
      \mathfrak{Q} = \mathfrak{Q}_{L} + \mathfrak{Q}_{R}
\end{equation}
with
\begin{equation}
\mathfrak{Q}_{L}=\sum_n\mathrm{Tr}\left(\tilde{\mathds D}_0[\sigma^z_n]\rho\right),\qquad \mathfrak{Q}_{R}=\sum_n\mathrm{Tr}\left(\tilde{\mathds D}_N[\sigma^z_n]\rho \right).
\end{equation}
Since $\rho_\infty$ is independent of time, it follows from equation \eqref{eq:spin-current-1} that the spin flow $\mathfrak{Q}$ in the non-equilibrium steady state vanishes.
Consequently, the contributions from each bath are inversely related, $\mathfrak{Q}_{L}  = -  \mathfrak{Q}_{R}$, indicating a spin current flowing from one side of the chain to the other. Applying formula \eqref{eq:dissipRL} and specializing to the case $\rho = \rho_\infty$, one can show that the spin flow $\mathfrak{Q}_{L}$ coming from the left bath is given by
\begin{widetext}
    \begin{equation}
        {\mathfrak{Q}}_{L}={2\lambda^2}
        \sum_k\frac{\phi_0(x_k)^2\phi_N(x_k)^2\left(C_{0,k}\tilde{C}_{N,k}-\tilde{C}_{0,k}C_{N,k}\right)}{\phi_0(x_k)^2C_{0,k} + \phi_N(x_k)^2\tilde{C}_{N,k}+\phi_0(x_k)^2\tilde{C}_{0,k} + \phi_N(x_k)^2 C_{N,k}}.
    \end{equation}
    
    A similar approach can be applied to study the heat flow $\mathfrak{h}(t)\coloneqq\frac{\mathrm d}{\mathrm dt}\mean{\Hamil\rho}$ in the non-equilibrium steady state 
    $\rho_\infty$, along with the associated thermal currents 
    $\mathfrak{h}_{L}$ and $\mathfrak{h}_{R}$ due to each bath. An analogous derivation shows that 
    $\mathfrak{h}_{L}  = - \mathfrak{h}_{R}$
    , and
    \begin{equation}
        \mathfrak{h}_{L}=\lambda^2
        \sum_k \frac{(x_k + \Delta) \phi_0(x_k)^2\phi_N(x_k)^2\left(C_{0,k}\tilde{C}_{N,k}-\tilde{C}_{0,k}C_{N,k}\right)}{\phi_0(x_k)^2C_{0,k} + \phi_N(x_k)^2\tilde{C}_{N,k}+\phi_0(x_k)^2\tilde{C}_{0,k} + \phi_N(x_k)^2 C_{N,k}}.
    \end{equation}
\end{widetext}
The smearing functions $h_\alpha$ appear in the interaction Hamiltonian of equation \eqref{eq:interaction-hamiltonian}, modulating the coupling strengths between each mode in the bath and the ends of the chain and allowing the introduction of a possible cutoff. In cases where the cutoff for the Bosonic fields has little impact on the physics at the energy scale of the spin chain, the smearing functions should vary slowly at this energy scale, allowing us to set $h_0(\omega)=h_N(\omega)=h$. Then, the average stationary spin and thermal flow simplify to
\begin{align}\label{eq:general-left-spin}
    &{\mathfrak{Q}}_{L}=4\pi h^2\lambda^2\times \nonumber \\
    &\sum_k \medmath{\frac{\phi_0(x_k)^2\phi_N(x_k)^2(n_0(x_k + \Delta)-n_N(x_k + \Delta))}{\phi_0(x_k)^2(2n_0(x_k + \Delta)+1)+\phi_N(x_k)^2(2n_N(x_k + \Delta)+1)}}
\end{align}
and
\begin{align}
    &\mathfrak{h}_{L}=2\pi h^2\lambda^2\times \nonumber \\
    \label{eq:general-left-heat}
    &\sum_k \medmath{\frac{(x_k + \Delta)\phi_0(x_k)^2\phi_N(x_k)^2(n_0(x_k + \Delta)-n_N(x_k + \Delta))}{\phi_0(x_k)^2(2n_0(x_k + \Delta)+1)+\phi_N(x_k)^2(2n_N(x_k + \Delta)+1)}}
\end{align}
These expressions simplify further for spin chains symmetric about their center, i.e. when
 $J_i = J_{N-1-i}$ and $B_i = B_{N-i}$. In this case, the special functions diagonalizing the single-particle Hamiltonian  $\mathrm{H}$ defined in \eqref{eq:hamiltonian-matrix} verify $\phi_0(x_k)^2 = \phi_N(x_k)^2$. The spin and heat flows can thus be expressed in terms of the first entry of the $(N+1) \times (N+1)$ matrix $\mathrm H + \Delta$, 
\begin{align}
    {\mathfrak{Q}}_{L}&= {2\pi\lambda^2}\bra{0}\frac{\sinh\left(\frac{\beta_N-\beta_0}{2}(\mathrm H + \Delta)\right)}{\sinh\left(\frac{\beta_0+\beta_N}{2}(\mathrm H + \Delta)\right)}\ket{0}, \nonumber\\
    \label{eq:fl1}
    \mathfrak{h}_{L}&=\pi \lambda^2\bra{0}(\mathrm H + \Delta)\frac{\sinh\left(\frac{\beta_N-\beta_0}{2}(\mathrm H+\Delta)\right)}{\sinh\left(\frac{\beta_0+\beta_N}{2}(\mathrm H + \Delta)\right)}\ket{0}.
\end{align}

For a general non-symmetric chain, this simplication is not possible, but one can still use the triangle inequality to derive simple upper bound on the spin and heat flow,

\begin{align}
{\mathfrak{Q}}_{L}\leq {2\pi h^2\lambda^2}
    \sum_k&\left|\phi_0(x_k)\phi_N(x_k)\right| \times \nonumber\\
    \label{eq:ineq1}
    &\frac{ \sinh\left(\frac{\beta_N-\beta_0}{2}(x_k + \Delta)\right)}{\sqrt{\sinh\left(\beta_N (x_k + \Delta)\right)\sinh\left(\beta_0 (x_k + \Delta)\right)}},
\end{align}

\begin{align}
    \mathfrak{h}_{L}\leq \pi h^2\lambda^2
    \sum_k &\left|\phi_0(x_k)\phi_N(x_k)\right|\times \nonumber \\
    \label{eq:ineq2}
    &\frac{(x_k + \Delta) \sinh\left(\frac{\beta_N-\beta_0}{2}(x_k + \Delta)\right)}{\sqrt{\sinh\left(\beta_N (x_k + \Delta)\right)\sinh\left(\beta_0 (x_k + \Delta)\right)}}.
\end{align}

Note that when the temperature gap between the two baths is small, the left and right terms in these relations are equal at leading order for a mirror-symmetric chain, demonstrating that the bound is strict and that chains symmetric with respect to their center exhibit optimal conductivity in the small temperature gap limit. 

By applying the triangle inequality a second time, one can further derive a less strict bound, which has the advantage of also being expressed in terms of the entries of a matrix,
\begin{widetext}
    \begin{equation}
    \begin{split}
        {\mathfrak{Q}}_{L} &\leq {\pi h^2\lambda^2}
       \bra{0} \frac{\sinh\left(\frac{\beta_N-\beta_0}{2}(\mathrm H+\Delta)\right)}{\sqrt{\sinh\left(\beta_N(\mathrm H+\Delta)\right)\sinh\left(\beta_0(\mathrm H+\Delta)\right)}}\ket{0}+ {\pi h^2\lambda^2}\bra{N}\frac{\sinh\left(\frac{\beta_N-\beta_0}{2}(\mathrm H+\Delta)\right)}{\sqrt{\sinh\left(\beta_N(\mathrm H+\Delta)\right)\sinh\left(\beta_0(\mathrm H+\Delta)\right)}} \ket{N},
       \end{split}
    \end{equation}
    \begin{equation}
    \begin{split}
        \mathfrak{h}_{L}&\leq \frac{\pi h^2\lambda^2}{2}
       \bra{0} \frac{ (\mathrm H + \Delta)\sinh\left(\frac{\beta_N-\beta_0}{2}(\mathrm H + \Delta)\right)}{\sqrt{\sinh\left(\beta_N(\mathrm H + \Delta)\right)\sinh\left(\beta_0(\mathrm H + \Delta)\right)}}\ket{0}+ \frac{\pi h^2\lambda^2}{2}\bra{N}\frac{(\mathrm H + \Delta)\sinh\left(\frac{\beta_N-\beta_0}{2}(\mathrm H + \Delta)\right)}{\sqrt{\sinh\left(\beta_N(\mathrm H + \Delta)\right)\sinh\left(\beta_0(\mathrm H + \Delta)\right)}} \ket{N}.
       \end{split}
    \end{equation}
\end{widetext}
In the following subsection, we apply these formulas and bounds to study how the spin and heat flow are affected by the temperature 
$T$ and the size of the system $N$, when the temperature difference between the baths is small,

\subsection{Small difference in temperature $|T_0 - T_N|$}\label{sec:4a}

According to Fourier's law, the heat flow $q := \mathfrak{h}_{L}$ is related to the temperature gradient $\nabla T$ through a linear relation
\begin{equation}
    q = - \kappa \nabla T,
\end{equation}
where $\kappa$ is a constant defined as the thermal conductivity. Heat diffusion is said to be \textit{anomalous} when $\kappa$ has a dependency on the size of the system \cite{liu2014anomalous}. Here, we compute the thermal conductivity $\kappa$ for mirror symmetric $XX$ spin chain in the limit where the gap in temperature is small. We further investigate how the temperature and inhomogeneities affect the type of anomalous transport it exhibits.  

In terms of the mean temperature $(T_0 + T_N)/2 = T $ (which approximate the temperature in the chain) and the temperature gap between the bath $(T_0 - T_N)  = \delta T $, one finds that the heat flow of equation \eqref{eq:fl1} is rewritten as
\begin{equation}\label{eq:heatfl}
    \mathfrak{h}_{L}=\pi \lambda^2 h^2\bra{0}(\mathrm H + \Delta)\frac{\sinh\left(\frac{2 \delta T}{4 T^2 - \delta T^2}(\mathrm H + \Delta)\right)}{\sinh\left(\frac{4 T}{4 T^2 - \delta T^2}(\mathrm H + \Delta)\right)}\ket{0}.  
\end{equation}
Thus, for a small temperature gap $|\delta T | \ll T$, the matrix arising in formula \eqref{eq:heatfl} reduces to
\begin{equation}
    \frac{\sinh\left(\frac{2 \delta T}{4 T^2 - \delta T^2}(\mathrm H + \Delta)\right)}{\sinh\left(\frac{4T}{4 T^2 - \delta T^2}(\mathrm H + \Delta)\right)}  = \frac{\delta T (\mathrm H + \Delta)}{2 T^2\sinh\left(\frac{(\mathrm H + \Delta)}{T}\right)}  + o(\delta T).
\end{equation}
Further making the identification $ \delta T= - N \nabla T$, the right-hand side of \eqref{eq:fl1} reduces to Fourier's law with a thermal conductivity given by
\begin{equation}\label{eq:cond1}
    \kappa = \frac{\pi \lambda^2 h^2 N}{2T^2} \bra{0} \frac{(\mathrm H + \Delta)^2}{\sinh\left(\frac{\mathrm H + \Delta}{T}\right)}\ket{0}
\end{equation}
In particular, it is possible to determine the relation between the conductivity and the temperature in regime of low or high temperature 

\begin{equation}\label{eq:cond2}
      \kappa \sim \frac{ \pi \lambda^2 h^2 N }{T^2} \bra{0} {(\mathrm H + \Delta)^2} e^{- \frac{\mathrm H + \Delta}{T}}\ket{0}, \quad \quad T \ll 1,
\end{equation}
\begin{equation} \label{eq:high-temp-reg}
      \kappa \sim \frac{ \pi \lambda^2 h^2 N }{2T} (B_0 + \Delta), \quad \quad T \gg 1.
\end{equation}

In the high-temperature regime, conductivity is inversely proportional to temperature as most bosons have energies exceeding the energy spectrum of the spin chain. It is interesting to note that at this range of temperature the magnetic field $B_0 = B_N$ at the ends of the chain is the only parameter influencing conductivity and that $\kappa \propto N$, indicating anomalous heat transport and ballistic transport for any (mirror symmetric) inhomogeneous $XX$ spin chain. The case of non-mirror symmetric chain in this regime is discussed in the following subsection.

In the low-temperature regime, conductivity decreases exponentially as the temperature is reduced. This is because the spin chains considered are gapped, where at low temperatures, the thermal bath lacks the necessary energy to create magnetic excitations in the spin chain. Equation \eqref{eq:cond1} shows that heat conductivity at low temperature can be influenced both by the single particle excitation energies (i.e. spectrum of $\mathrm H$) and the overlaps $\phi_0(x_k) = \phi_N(x_k)$ of the associated modes with local excitation at the ends of the chain. Therefore, different type of transport can be observed depending on the inhomogeneous couplings $J_i$ and local magnetic fields $B_i$. 

As example, we compare different chains whose single excitation energy spectrum $x_k + \Delta$ are scaled to be from $E_{min}$ to $E_{max}$. For the homogeneous $XX$ spin chain in the $N \rightarrow \infty$ and $T \ll E_{min}$ limit, one finds ballistic transport ($\kappa \propto N$),
\begin{equation}
    \kappa \approx \frac{ 4 \sqrt{\pi} \lambda^2 h^2  E_{min}^2  e^{-E_{min}/T}}{ \sqrt{T}(E_{max}-E_{min})^2 } N
\end{equation}
In contrast, for the Krawtchouk chain with 
$p = 1/2$, we can compute \eqref{eq:cond1} exactly and observe a conductivity decreases with the size of the chain, indicating sub-diffusive transport at leading order,
\begin{equation}
      \kappa \approx  \frac{ \pi \lambda^2 h^2  E_{min}^2  N}{ T^22^N } e^{- E_{min}/T}.
\end{equation}
This indicates that a spin chain's ability to achieve perfect state transfer does not guarantee efficient scaling of heat flow at low temperatures. Indeed, the homogeneous chain, which does not exhibit perfect state transfer, shows ballistic transport, whereas the Krawtchouk chain, which allows for perfect state transfer, exhibits sub-diffusive transport in this regime. 

An explanation for the difference between the two chains lie in the probability $|\phi_0(x_k )|^2 = |\phi_N(x_k)|^2$ of an excitation of energy $x_k + \Delta$ in the chain to visit either ends. This quantity is much smaller in the Krawtchouk case for most energies. Indeed, for the Krawtchouk case we have for $k = 0, 1, \dots, N$,
\begin{equation}
    |\phi_0(x_k)|^2 = \binom{N}{k}\frac{1}{2^N}, \quad x_k + \Delta = (E_{max} - E_{min})\frac{k}{N} + E_{min}
\end{equation}
This is exponentially vanishing as $N$ increases, except for $k \approx N/2$ and $x_k + \Delta \approx \frac{E_{max} + E_{min}}{2}$. In the homogeneous chain, the probability of each mode to reach the ends decreases much more slowly as 
$1/N$,
\begin{equation}
    |\phi_0(x_k)|^2 = \frac{8(E_{max}-x_k - \Delta) (x_k+ \Delta -E_{min})}{(N+1)(E_{max}-E_{min})^2}.
\end{equation}
At low temperatures, when only the lowest energy modes are excited, these modes can still facilitate heat transport. 
\subsection{Ballistic to subdiffusive transition}\label{sec:4b}

Formulas \eqref{eq:high-temp-reg} shows that any inhomogeneous chain with mirror symmetric couplings $J_i$ and local magnetic field $B_i$ exhibit ballistic transport at high temperature. We now show that this property can be highly sensible to the couplings in the chain and vanish once the symmetry is broken. In the high temperature and small temperature gap limit, the inequality \eqref{eq:ineq2} reduces to
\begin{equation}\label{eq:ineq2-mod-2}
    \mathfrak{h}_{L}\leq \frac{\pi h^2\lambda^2 \delta T}{T} M(\mathrm H + \Delta),
\end{equation}
where
\begin{equation}\label{eq:ineq2-mod}
   M(\mathrm H + \Delta)=
    \sum_k |\phi_0(x_k)\phi_N(x_k)|(x_k + \Delta).
\end{equation}
This implies the following upper bound on the conductivity $\kappa$,
\begin{equation}
    \kappa \leq \frac{2\pi h^2\lambda^2 }{T} M(\mathrm H + \Delta),
\end{equation}
Note that the coefficient $M(\mathrm H + \Delta)$ is the only part that depends on the size of the chain $N$ and the choice of parameters $J_i$ and $B_i$. In the case where this coefficient decreases with $N$, we can thus deduce that the transport is subdiffusive. First, we consider this quantity for homogeneous and Krawtchouk chain ($p =1/2$) perturbed by a magnetic field whose strength is linear in the position $i$ in the chain, i.e.
\begin{equation}
    B_i \rightarrow B_i + \xi \frac{ i}{N},
\end{equation}
where $\xi$ is the strength of the perturbation.
\begin{figure}
\begin{subfigure}[t]{0.45\textwidth}
  \centering
  \includegraphics[width=.8\linewidth]{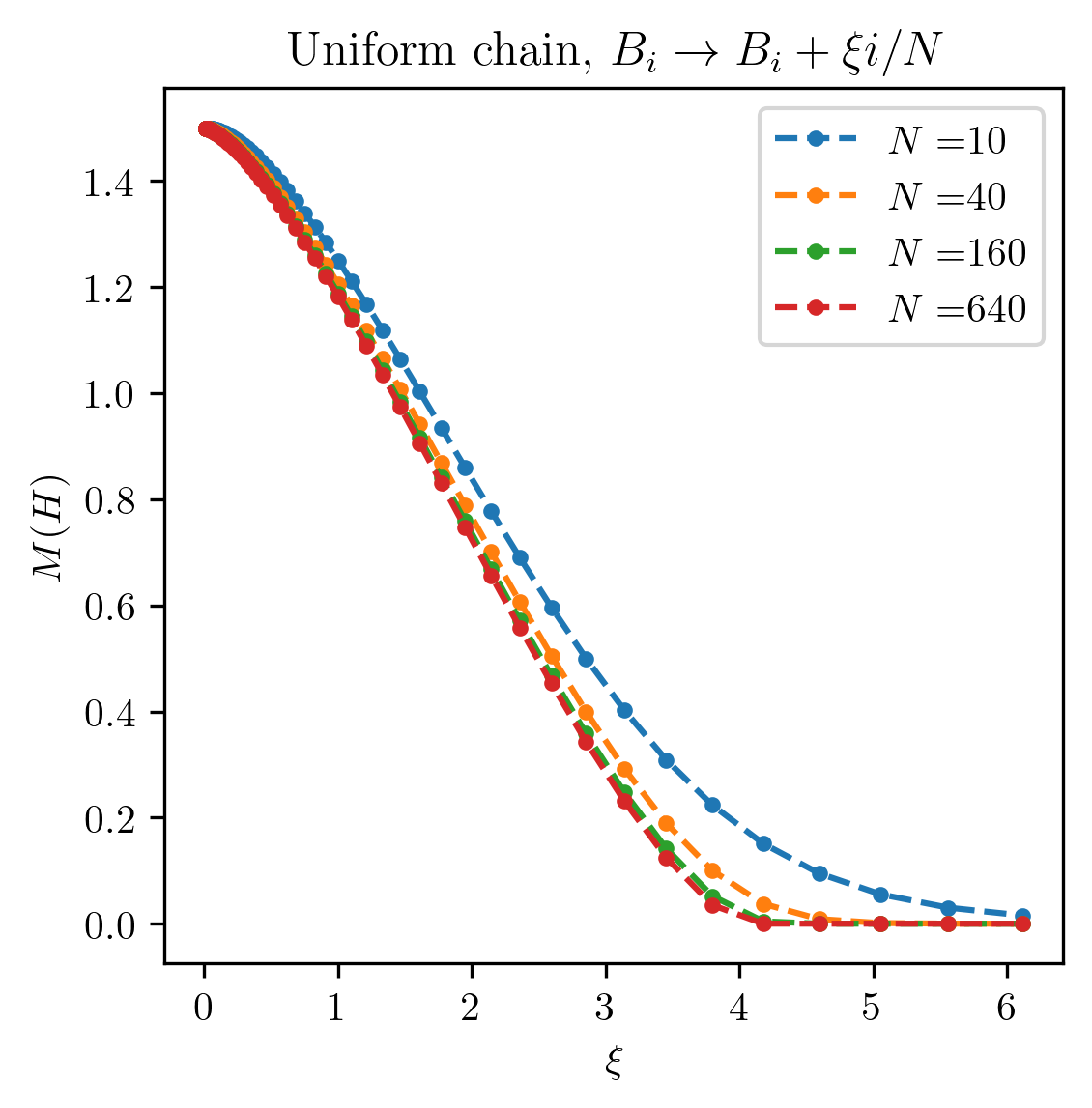}
  \label{fig:sfig1}
\end{subfigure}\hfill
\begin{subfigure}[t]{0.45\textwidth}
  \centering
  \includegraphics[width=.8\linewidth]{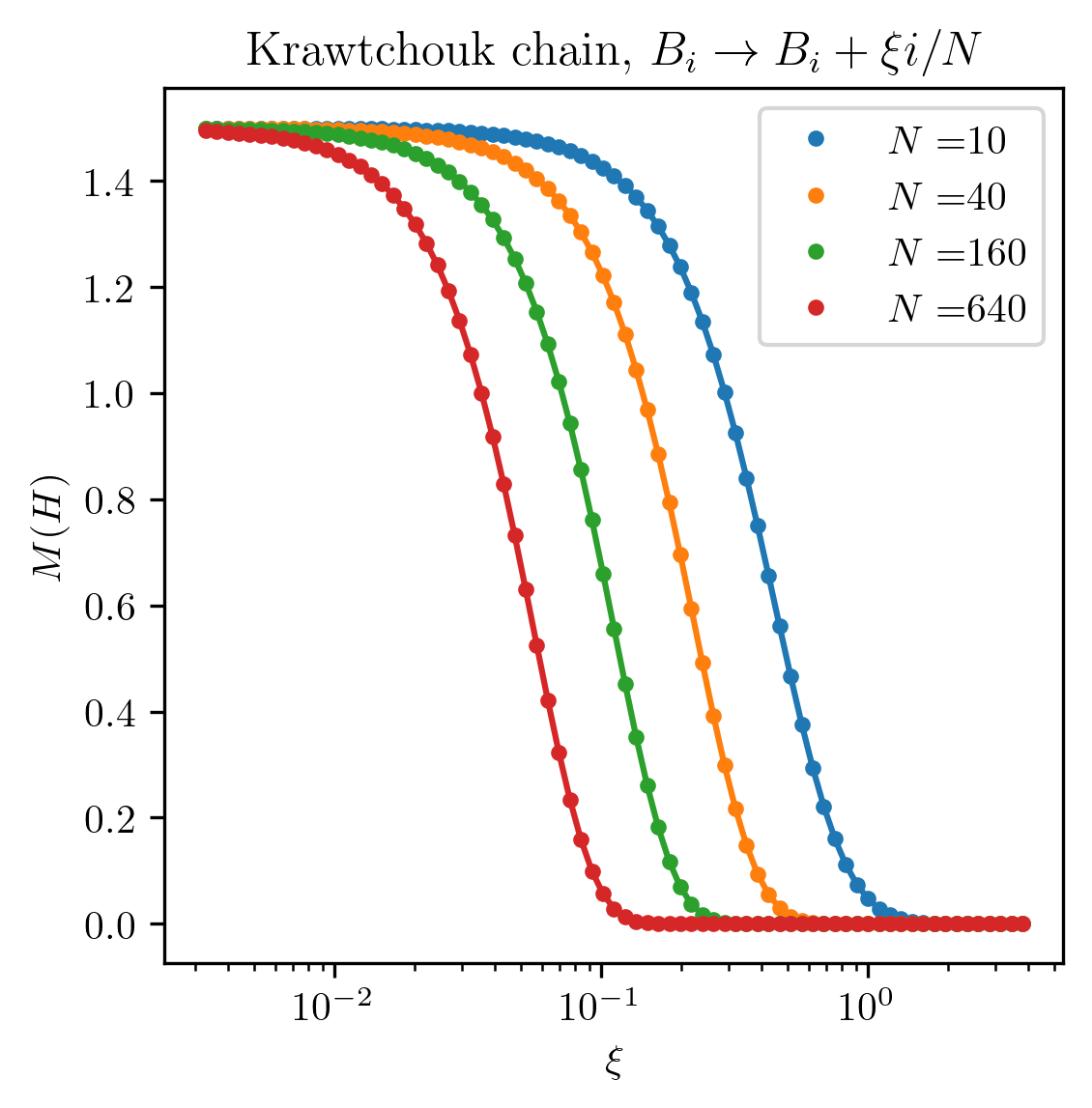}
  \label{fig:sfig2}
\end{subfigure}
\caption{Coefficient $M(\mathrm{H} + \Delta)$ arising in the upper bound for the heat flow of homogeneous and Krawtchouk (with $p = 1/2$) chains perturbed by a linear magnetic field $\xi \frac{i}{N}$. The lines in the bottom figure correspond to the analytical predictions.}
\label{fig:lin}
\end{figure}

\begin{figure}
\begin{subfigure}[t]{0.45\textwidth}
  \centering
  \includegraphics[width=.8\linewidth]{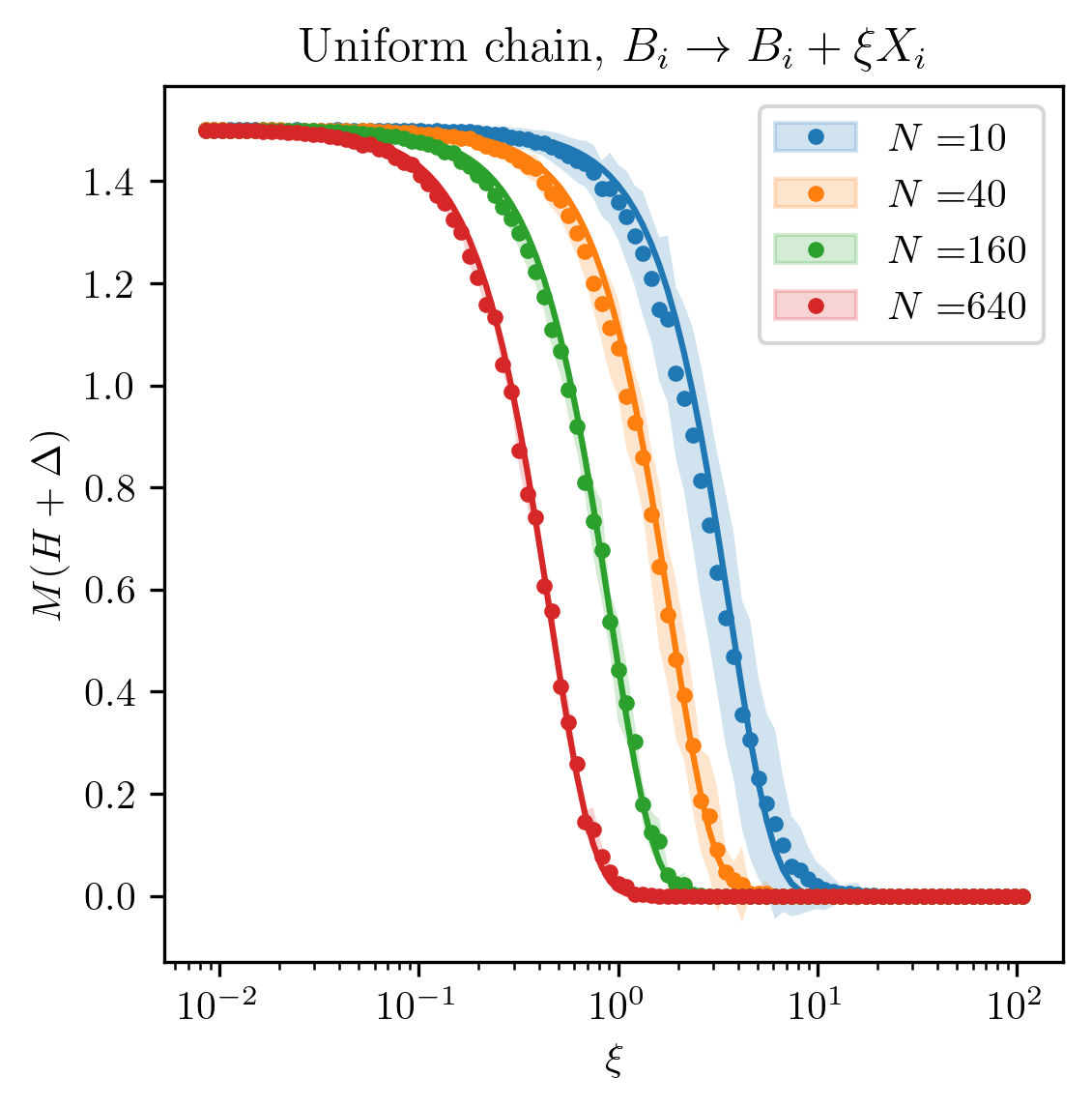}
  \label{fig:sfig3}
\end{subfigure}\hfill
\begin{subfigure}[t]{0.45\textwidth}
  \centering
  \includegraphics[width=.8\linewidth]{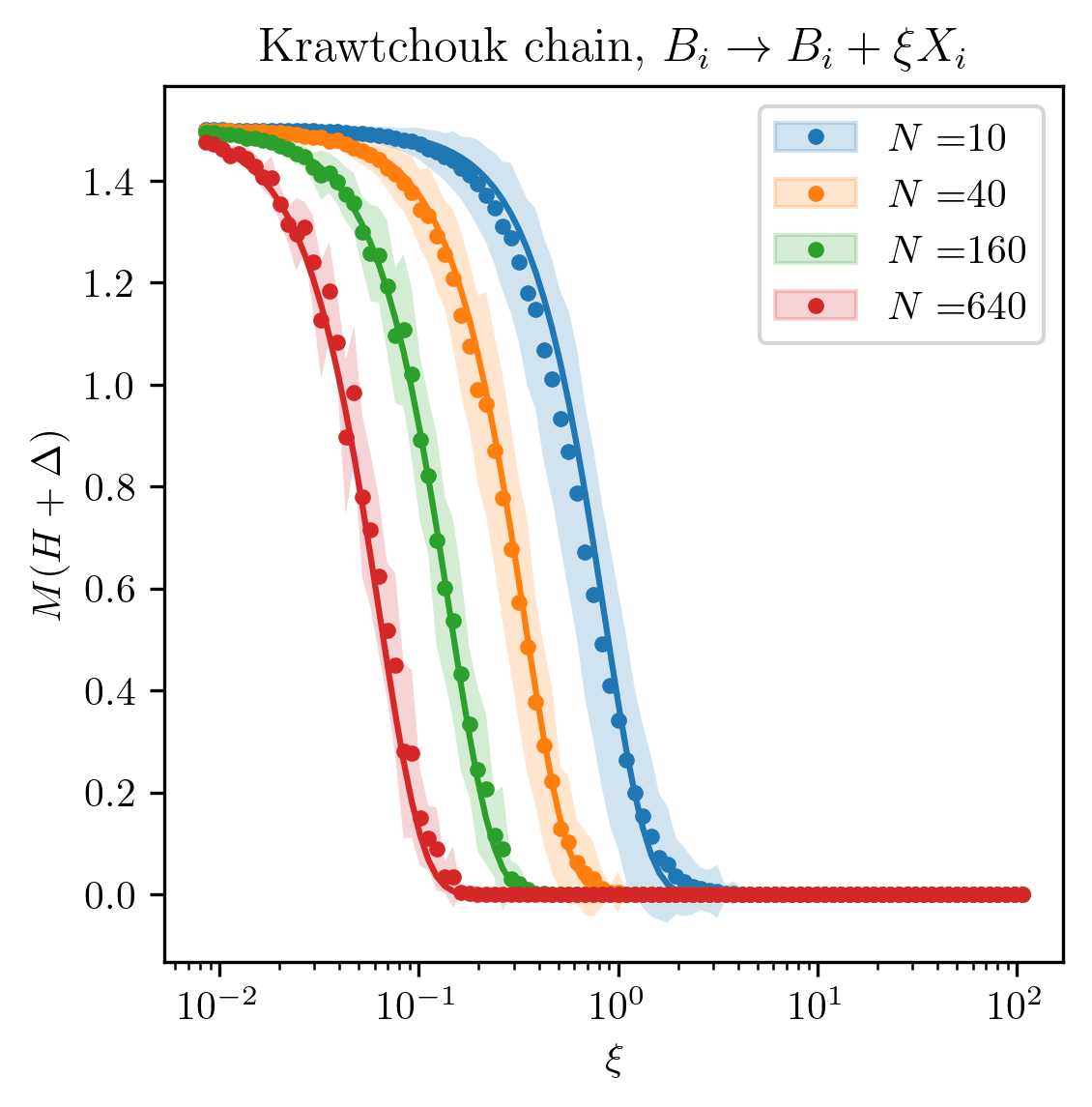}
  \label{fig:sfig4}
\end{subfigure}
\caption{Average coefficient $M(\mathrm{H} + \Delta)$ arising in the upper bound for the heat flow of homogeneous and Krawtchouk (with $p = 1/2$) chains perturbed by a random magnetic field $\xi X_i$ with $X_i \sim U(0,1)$. The shaded areas represent the standard deviation. The lines were obtained by fitting the conjectures \eqref{eq:conj1} and \eqref{eq:conj2} for the uniform and Krawtchouk chains respectively. }
\label{fig:rand}
\end{figure}

For both cases, $M(\mathrm{H} + \Delta)$ can be computed numerically, as shown in Figure \ref{fig:lin}. For the Krawtchouk, one can further obtains an analytical results since introducing a linear magnetic field amounts to change the value of the parameter $p$,
\begin{align}
    \mathfrak{h}_{L}&\leq  \frac{\pi h^2\lambda^2  (E_{max} + E_{min})\delta T}{2 T} (2 \sqrt{p(1-p)})^N  \nonumber\\
    \label{eq:ineq3}
    &\qquad= \frac{\pi h^2\lambda^2 (E_{max} + E_{min})\delta T}{2 T\left({1+\xi^2/4}\right)^{N/2}} .
\end{align}
The heat flow vanishes exponentially as we scale the system ($N \rightarrow \infty$) for any strength $\xi$ of the perturbation. In contrast, we see in Figure \ref{fig:lin} that the homogeneous chain is much less sensible, with transport that decrease with $\xi$ but with little impact of $N$. This does not mean that the ballisitic transport in the homogeneous chain is stable under any perturbation, as it appears that both chains are fragile under random perturbation $X_i$ of the magnetic field taken from the uniform distribution $U(0,1)$ on the interval $[0,1]$,
\begin{equation}
    B_i \rightarrow B_i + \xi X_i, \quad X_i \sim U(0,1).
\end{equation}
Indeed, the coefficient $M(\mathrm{H} + \Delta)$ then seems to decrease exponentially with $N$ for both chain and for any $\xi$, as illustrated in Figure \ref{fig:rand}.
For the uniform chain and the Krawtchouk chain, we conjecture based on numerical results that
\begin{equation}\label{eq:conj1}
 M(\mathrm{H} + \Delta) \propto e^{-\xi^2 N},
\end{equation}
and
\begin{equation}\label{eq:conj2}
     M(\mathrm{H} + \Delta) \propto e^{-\xi^2 N \log N},
\end{equation}
indicating subdiffusive transport.

\subsection{High temperature gap regime}\label{sec:4c}
\begin{figure}
\begin{subfigure}[t]{0.45\textwidth}
  \centering
  \includegraphics[width=\linewidth]{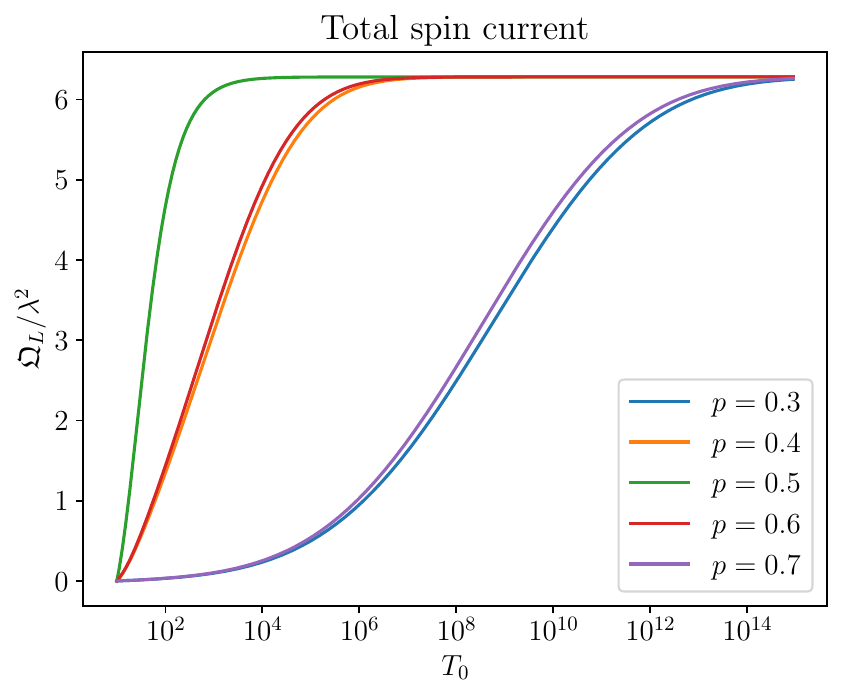}
  \label{fig:spin-current}
\end{subfigure}\hfill
\begin{subfigure}[t]{0.45\textwidth}
  \centering
  \includegraphics[width=\linewidth]{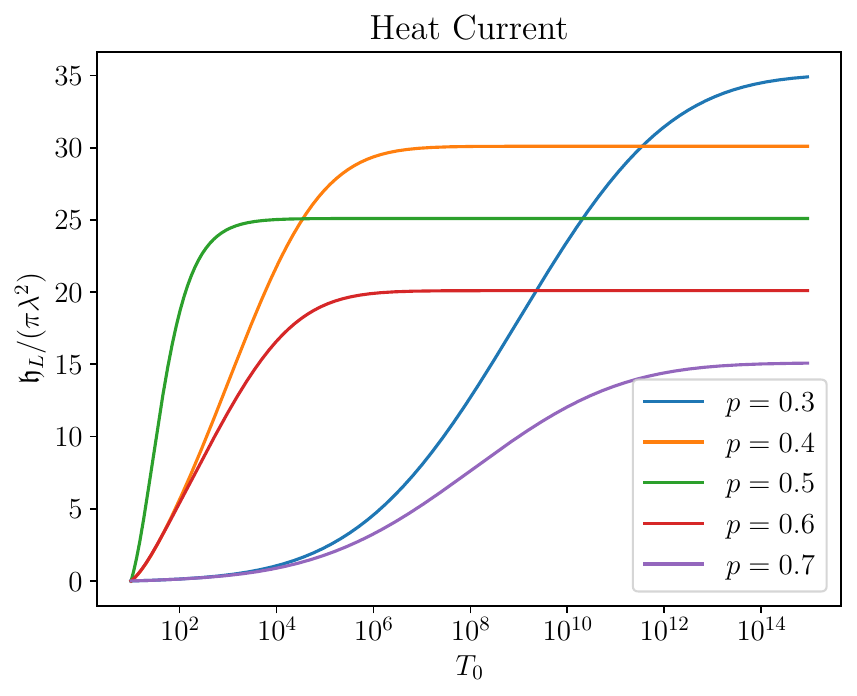}
  \label{fig:heat-current}
\end{subfigure}
\caption{Current strengths in various Krawtchouk chains. The right bath is kept at fixed temperature $T_N=10$. The system has a background magnetic field $\Delta=0.1$ and $N=50$ chain sites.}
\label{fig:high-temperature-gap-regime}
\end{figure}

When the two baths have very different temperatures (say $\beta_0\ll\beta_N$), the spin and heat currents simplify greatly. In \eqref{eq:general-left-spin} and \eqref{eq:general-left-heat}, $n_0(x_k + \Delta)$ will be dominant and taking the limit $\beta_0\to0$, we are left with
\begin{equation}
    \mathfrak{Q}_L=2\pi\lambda^2 \qquad \mathfrak{h}_L=\pi\lambda^2 (B_N + \Delta) \ .
\end{equation}
This and the monotonicity of the currents with respect to bath temperatures means that the currents never go past these strengths as $T_0$ increases. This is illustrated for the Krawtchouk chain in figure \ref{fig:high-temperature-gap-regime}. For this example, only the rate of convergence seems to differ between different values of the parameter $p$ with $p=1/2$ having the fastest convergence.

This result is independent of the choice of $J_i$ and $B_i$ for $ i = 0, 1, \dots, N-1$, indicating that any inhomogeneities have minimal impact on the currents in this regime. This likely arises because the bath at high temperature $T_0$ is saturating every modes within the chain. Consequently, the amount of energy leaving the system is simply proportional to $B_N + \Delta$, which is the energy associated with a spin-up at the site coupled to the exiting bath.

\section{Discussion}

We studied spin chains with arbitrary nearest-neighbor couplings with their extremities in heat baths. An exact expression for the density matrix of the non-equilibrium steady state was found in terms of the spectrum and stationary state wavefunctions of the single-particle Hamiltonian. Similar results were derived for the associated spin and heat currents. The scaling of these currents with the size of the chains was investigated in different regimes dictated by the mean temperature and the temperature difference of the baths. The low-temperature regime was identified as one where the inhomogeneities have a significant impact, with ballistic or diffusive transport depending on the wavefunction of the stationary states. At high temperatures, it was observed that chains with symmetric couplings and magnetic fields with respect to the middle enabled ballistic transport. This property was shown to be easily lost in non mirror-symmetric chains. 

While one might have expected to observe higher currents in chains with perfect state transfer, this was not found to be the case. The homogeneous chain, which does not allow perfect state transfer, had a low-temperature thermal conductivity scaling with 
$N$, whereas the Krawtchouk chain with 
$p=1/2$, which exhibits perfect state transfer, had a thermal conductivity that vanishes exponentially with the size of the system in the same regime. Still, it is interesting to note that both perfect state transfer and ballistic transport at high temperatures require the chain to be mirror symmetric.
\section*{Acknowledgements}

PAB hold an Alexander-Graham-Bell scholarship from the Natural Sciences and Engineering Research Council of Canada (NSERC). The research of LV is supported by a Discovery Grant from NSERC.

\appendix*
\section{Derivation of the Lindblad master equation}\label{sec:appendixA}

For sake of completeness we collect in this Appendix the derivation
of the Lindblad master equation governing the reduced dynamics of an inhomogeneous spin chain weakly interacting with thermal baths (for general references on this subject, see \cite{BENATTI_2005,Rivas_2012,Alicki2007}.)

We consider a composite system whose Hamiltonian is given by
\[\Hamil_{\mathrm{tot}}=\Hamil+\Hamil_\text{env}+\lambda \Hamil_I\]
where $\Hamil$ is the Hamiltonian of the system we are interested in, $\Hamil_E$ is the Hamiltonian of the environment in which evolves the system and $\Hamil_I$ is an Hamiltonian describing the interactions between the system and the environment. The interaction Hamiltonian can be decomposed as
\begin{equation}
    \Hamil_I=\sum_i E_i\otimes S_i
\end{equation}
with 
\begin{equation}
    \mean{E_i}=\mathrm{Tr}_\text{env}(E_i)=0
\end{equation}
This choice is justified because these mean values were non-zero, one could always absorb those in the system's Hamiltonian by defining
\begin{equation}
    \Hamil'=\Hamil+\sum_i\mean{E_i}S_i\ .
\end{equation}
\begin{leftrule}
    For the open spin chain system this means that
    \begin{gather}
        \begin{gathered}
            E_0=\int_0^\infty\!\mathrm{d}\nu\ h_0(\nu)\mathfrak{a}_0(\nu) \\
            E_1=\int_0^\infty\!\mathrm{d}\nu\ h_N(\nu)\mathfrak{a}_N(\nu) \\
            E_2=E_0^\dagger=\int_0^\infty\!\mathrm{d}\nu\ h_0(\nu)\mathfrak{a}_0^\dagger(\nu) \\
            E_3=E_1^\dagger=\int_0^\infty\!\mathrm{d}\nu\ h_N(\nu)\mathfrak{a}_N^\dagger(\nu)
        \end{gathered}
    \end{gather}
    and
    \begin{gather}
        \begin{gathered}
            S_0=\sigma_0^+ \\
            S_1=\sigma_N^+ \\
            S_2=S_0^\dagger=\sigma_0^- \\
            S_3=S_1^\dagger=\sigma_N^- \ .
        \end{gathered}
    \end{gather}
\end{leftrule}

Starting with the von Neumann equation
\begin{equation}
    \dot\rho_\text{tot}(t)=-i\left[\Hamil_\text{tot},\rho_\text{tot}(t)\right],
\end{equation}
one can go to the interaction picture where interaction operators $\hat O$ take the form
\begin{equation}
    \hat O(t)=e^{i(H+H_E)t}Oe^{-i(H+H_E)t}
\end{equation}
In this picture, the von Neumann equation reads
\begin{equation}
    \dot{\hat\rho}_\mathrm{tot}(t)=-i\lambda\left[\hat \Hamil_I(t),\hat\rho_\mathrm{tot}(t)\right]\ .
\end{equation}
This can be solved as a perturbative series in $\lambda$
\begin{align}
    \dot{\hat\rho}_\mathrm{tot}(t)&=-i\lambda\left[\hat \Hamil_I(t),\hat\rho_\mathrm{tot}(0)\right] \nonumber \\
    &\quad-\lambda^2\int_{0}^{t}\mathrm d s\left[\hat \Hamil_I(t),\left[\hat \Hamil_I(s),\hat\rho_\mathrm{tot}(t)\right]\right] +O(\lambda^3)\ .
\end{align}
Assuming weak coupling, higher order terms can be discarded. Tracing over the environment, this becomes an equation for the evolution of the reduced density matrix of the system
\begin{align}
    \dot{\hat\rho}(t)&=-i\lambda\mathrm{Tr}_E\left[\hat \Hamil_I(t),\hat\rho_\mathrm{tot}(0)\right] \nonumber \\
    &\quad-\lambda^2\int_{0}^{t}\mathrm d s\mathrm{Tr}_E\left[\hat \Hamil_I(t),\left[\hat \Hamil_I(s),\hat\rho_\mathrm{tot}(t)\right]\right]
\end{align}
The condition $\mean{E_i}=0,\forall i$ makes the first term vanish. The only term left is the quadratic term which can be rewritten as
\begin{align}
    \dot{\hat\rho}(t)=-\lambda^2\int_{0}^{t}\mathrm d s\mathrm{Tr}_E&\left(\hat \Hamil_I(t)\hat \Hamil_I(t-s)\hat\rho_\mathrm{tot}(t)\right.\nonumber \\
    &\quad -\left.\hat \Hamil_I(t-s)\hat\rho_\mathrm{tot}(t)\hat \Hamil_I(t)\right)+h.c.
\end{align}
To progress further, we decompose the system operators $S_i$ in term of transition energies $\omega=\varepsilon'-\varepsilon$ for the system
\begin{equation}\label{eq:lindblad-ops-def}
    S_i(\omega)=\sum_{\varepsilon}\Pi(\varepsilon)S_i\Pi(\varepsilon+\omega)\ .
\end{equation}
where $\Pi(\varepsilon)$ is the projector on the eigenspace of eigenvalue $\varepsilon$ of the system Hamiltonian $H$. These $S_i(\omega)$ are what is called the Lindblad operators. It is readily seen that
\begin{equation}
    S_i=\sum_{\omega}S_i(\omega)\ .
\end{equation}
\begin{leftrule}
    Using the definition \eqref{eq:lindblad-ops-def}, we have
    \begin{equation}
        S_0(\omega)=\sum_{\epsilon}\Pi(\epsilon)\sigma_0^+\Pi(\epsilon+\omega)\ .
    \end{equation}
    Performing an inverse Jordan-Wigner transformation, using the eigenbasis $\doubleket{\vec q}$ and going to the diagonal fermionic operators, this becomes
    \begin{equation}
        S_0(\omega)=\sum_{k=0}^N\phi_0(x_k)\sum_{\vec{q},\vec{q}'|E_{\vec q'}-E_{\vec q}=\omega}\doubleketbra{\vec q}{\vec q}b_k^\dagger\doubleketbra{\vec q'}{\vec q'}
    \end{equation}
    Clearly, it must be that $\doubleket{\vec q}=b_k^\dagger\doubleket{\vec q'}$, therefore the only allowed transition energies are the single mode energies $E_k$. The only non zero cases are then
    \begin{align}
        S_0(-E_k)&=\phi_0(x_k)b_k^\dagger\sum_{\vec{q}'}\doubleketbra{\vec q'}{\vec q'}b_kb_k^\dagger\doubleketbra{\vec q'}{\vec q'} \nonumber \\
        &=\phi_0(x_k)b_k^\dagger
    \end{align}
    
    Similarly one can find the other Lindblad operators which are
    \begin{gather}\label{eq:other-s}
        \begin{gathered}
            S_1(-E_k)=\phi_N(x_k)b_k^\dagger \\
            S_2(E_k)=(-1)^{\sum_{y=0}^Nb_y^\dagger b_y}\phi_0(x_k)b_k \\
            S_3(E_k)=(-1)^{\sum_{y=0}^Nb_y^\dagger b_y}\phi_N(x_k)b_k
        \end{gathered}
    \end{gather}
\end{leftrule}
In the interaction picture,
\begin{equation}
    \hat S_i(\omega,t)=e^{-i\omega t}S_i(\omega)\ .
\end{equation}
The interaction Hamiltonian can therefore be rewritten as
\begin{equation}\label{eq:AHint}
    \hat \Hamil_I(t)=\sum_{i,\omega}e^{-i\omega t}S_i(\omega)\otimes\hat E_i(t)=\sum_{i,\omega}e^{i\omega t}S_i^\dagger(\omega)\otimes\hat E_i^\dagger(t)
\end{equation}
where the last equality follow from the hermicity of $\hat H_I(t)$. The evolution for the reduced density can be then rewritten with the help of \eqref{eq:AHint} as
\begin{align}
    \dot{\hat \rho}(t)=-\lambda^2&\sum_{\omega,\omega',i,j}e^{i(\omega-\omega')t}\Gamma_{ij}(\omega',t)\times \nonumber \\
    &\left(S_i^\dagger(\omega) S_j(\omega')\hat\rho(t)-S_j(\omega')\hat\rho(t) S_i^\dagger(\omega)\right)+h.c.
\end{align}
where
\begin{equation}
    \Gamma_{ij}(\omega,t)=\int_0^t\mathrm{d}s\ e^{i\omega s}\mathrm{Tr}_E\left(\hat \rho_E(0)\hat E_i^\dagger(0)\hat E_j(-s)\right)\ .
\end{equation}
Given that the partial trace in this expression represents a correlation between bath operators at time $-s$ and $0$ and we expect no correlations between these operators at distant times, we can push the integration bound to infinity and define
\begin{equation}\label{eq:gamma}
    \Gamma_{ij}(\omega):= \lim_{t\to\infty}\Gamma_{ij}(\omega,t)=\!\int_0^\infty\!\mathrm{d}s\ e^{i\omega s}\mathrm{Tr}_E\left(\rho_EE_i^\dagger\hat E_j(-s)\right)
\end{equation}
such as to rewrite the evolution equation as
\begin{align}
    \dot{\hat \rho}(t)=-\lambda^2&\sum_{\omega,\omega',i,j}e^{i(\omega-\omega')t}\Gamma_{ij}(\omega')\times\nonumber\\
    &\left(S_i^\dagger(\omega) S_j(\omega')\hat\rho(t)-S_j(\omega')\hat\rho(t) S_i^\dagger(\omega)\right)+h.c.
\end{align}
\begin{leftrule}
    For the open spin chain, these quantities are computed to be
    \begin{gather}\label{eq:big-gamma}
        \begin{gathered}
            \Gamma_{00}(\omega)=\pi h_0(-\omega)^2n_0(-\omega)\Theta(-\omega) \\
            \qquad\qquad+i\mathrm{P.V.}\!\int_0^\infty\!\mathrm{d}\nu\ \frac{h_0^2(\nu)n_0(\nu)}{\omega+\nu} \\
            \Gamma_{11}(\omega)=\pi h_N(-\omega)^2n_N(-\omega)\Theta(-\omega) \\
            \qquad\qquad+i\mathrm{P.V.}\!\int_0^\infty\!\mathrm{d}\nu\ \frac{h_N^2(\nu)n_N(\nu)}{\omega+\nu} \\
            \Gamma_{22}(\omega)=\pi h_0(\omega)^2\left(n_0(\omega)+1\right)\Theta(\omega) \\
            \qquad\qquad\qquad+i \mathrm{P.V.}\!\int_0^\infty\!\mathrm{d}\nu\frac{h_0(\nu)^2\left(n_0(\omega)+1\right)}{\omega-\nu} \\
            \Gamma_{33}(\omega)=\pi h_N(\omega)^2\left(n_N(\omega)+1\right)\Theta(\omega) \\
            \qquad\qquad\qquad+i \mathrm{P.V.}\!\int_0^\infty\!\mathrm{d}\nu\frac{h_N(\nu)^2\left(n_N(\omega)+1\right)}{\omega-\nu}
        \end{gathered}
    \end{gather}
    and zero otherwise, where $n_\alpha(\nu)=\frac{1}{e^{\beta_\alpha \nu}-1}$ are bosonic mean thermal occupation numbers, $\Theta(\omega)$ is the Heavyside step-function and $\mathrm{P.V.}$ denotes the Cauchy principal value of the integral.
\end{leftrule}

Back to the dynamical equation, what is called the secular approximation, where only terms with $\omega=\omega'$ are kept, can be performed which gives
\begin{align}
    \dot{\hat \rho}(t)=-\lambda^2&\sum_{\omega,i,j}\Gamma_{ij}(\omega)\times\nonumber\\
    &\left(S_i^\dagger(\omega) S_j(\omega)\hat\rho(t)-S_j(\omega)\hat\rho(t) S_i^\dagger(\omega)\right)+h.c.
\end{align}
Renaming some summation indices and defining the quantities
\begin{equation}
    \begin{split}
        \pi_{ij}(\omega)&=\frac{-i}{2}\left(\Gamma_{ij}(\omega)-\Gamma_{ji}^*(\omega)\right) \\
        \gamma_{ij}&=\Gamma_{ij}(\omega)+\Gamma_{ji}^*(\omega)
    \end{split}
\end{equation}
the evolution equation takes the form
\begin{align}
        \dot{\hat \rho}(t)=&-\lambda^2i\sum_{\omega,i,j}\pi_{ij}(\omega)\left[S_i^\dagger(\omega)S_j(\omega),\hat\rho(t)\right] \nonumber \\
        &+\lambda^2\sum_{\omega,i,j}\gamma_{ij}(\omega)\times \nonumber \\
        &\quad\left(S_j(\omega)\hat\rho(t) S_i^\dagger(\omega)-\frac{1}{2} \left\{\hat\rho(t),S_i^\dagger(\omega)S_j (\omega)\right\}\right)
\end{align}
\begin{leftrule}
    It follows from \eqref{eq:big-gamma} that for the open spin chain, the coefficients are
    \begin{gather}
        \begin{gathered}\label{eq:small-gamma}
            \gamma_{00}(\omega)=2\pi h_0(-\omega)^2n_0(-\omega)\Theta(-\omega) \\
            \gamma_{11}(\omega)=2\pi h_N(-\omega)^2n_N(-\omega)\Theta(-\omega) \\
            \gamma_{22}(\omega)=2\pi h_0(\omega)^2(n_0(\omega)+1)\Theta(\omega) \\
            \gamma_{33}(\omega)=2\pi h_N(\omega)^2(n_N(\omega)+1)\Theta(\omega)
        \end{gathered}
    \end{gather}
    
    and
    
    \begin{gather}\label{eq:pi}
        \begin{gathered}
            \pi_{00}(\omega)=i\mathrm{P.V.}\!\int_0^\infty\!\mathrm{d}\nu\ \frac{h_0^2(\nu)n_0(\nu)}{\omega+\nu} \\
            \pi_{11}(\omega)=i\mathrm{P.V.}\!\int_0^\infty\!\mathrm{d}\nu\ \frac{h_N^2(\nu)n_N(\nu)}{\omega+\nu} \\
            \pi_{22}(\omega)=i \mathrm{P.V.}\!\int_0^\infty\!\mathrm{d}\nu\frac{h_0(\nu)^2(n_0(\nu)+1)}{\omega-\nu} \\
            \pi_{33}(\omega)=i \mathrm{P.V.}\!\int_0^\infty\!\mathrm{d}\nu\frac{h_N(\nu)^2(n_N(\nu)+1)}{\omega-\nu}\ .
        \end{gathered}
    \end{gather}
\end{leftrule}
If we bring back the evolution equation to the Schrödinger picture, it becomes what is called the Lindblad Master equation
\begin{equation} 
    \dot{\rho}(t)=-i\left[\Hamil+\Hamil_{LS},\rho\right]+\mathds{D}[\rho(t)]
\end{equation}
where
\begin{equation} \label{eq:lamb-shift}
    \Hamil_{LS}=\sum_{\omega,i,j}\pi_{ij}(\omega)S_i^\dagger(\omega)S_j(\omega)
\end{equation}
and
\begin{equation}\label{eq:dissipator}
    \mathds{D}[\rho]=\!\lambda^2\!\sum_{\omega,i,j}\gamma_{ij}(\omega)\!\left(\!S_j(\omega)\rho S_i^\dagger(\omega)-\frac{1}{2}\! \left\{\rho,S_i^\dagger(\omega)S_j (\omega)\right\}\!\right)
\end{equation}

\begin{widetext}
Therefore, the Lamb-shift Hamiltonian for the open spin chain is
\begin{equation}\label{eq:dirty-HLS}
        \Hamil_{LS}=\sum_{k=0}^N\sum_{\alpha\in\{0,N\}}\phi_\alpha(x_k)^2\left(\mathrm{P.V.}\!\int_0^\infty\!\mathrm{d}\nu\ \frac{h_\alpha^2(\nu)n_\alpha(\nu)}{\nu-E_k}b_kb_k^\dagger+\mathrm{P.V.}\!\int_0^\infty\!\mathrm{d}\nu\ \frac{h_\alpha^2(\nu)(n_\alpha(\nu)+1)}{E_k-\nu}b_k^\dagger b_k\right)\ .
\end{equation}
and the dissipative contribution is
\begin{gather}\label{eq:dirty-Diss}
    \begin{gathered}
        \mathds{D}[\rho]=2\pi\lambda^2\sum_{\alpha\in\{0,N\}}\sum_{k=0|E_k>0}^N\phi_\alpha(x_k)^2\left(h_\alpha(E_k)^2n_\alpha(E_k)\left(b_k^\dagger\rho(t)b_k-\frac{1}{2} \left\{\rho(t),b_xb_x^\dagger\right\}\right)\right. \\
        +\left.h_\alpha(E_k)^2(n_\alpha(E_k)+1)\left(b_k\rho(t)b_k^\dagger-\frac{1}{2} \left\{\rho(t),b_k^\dagger b_k\right\}\right)\right).
    \end{gathered}
\end{gather}    
\end{widetext}

\nocite{*}

\bibliography{bib_inhomogeneous_bosons}

\begin{thebibliography}{42}%
\makeatletter
\providecommand \@ifxundefined [1]{%
 \@ifx{#1\undefined}
}%
\providecommand \@ifnum [1]{%
 \ifnum #1\expandafter \@firstoftwo
 \else \expandafter \@secondoftwo
 \fi
}%
\providecommand \@ifx [1]{%
 \ifx #1\expandafter \@firstoftwo
 \else \expandafter \@secondoftwo
 \fi
}%
\providecommand \natexlab [1]{#1}%
\providecommand \enquote  [1]{``#1''}%
\providecommand \bibnamefont  [1]{#1}%
\providecommand \bibfnamefont [1]{#1}%
\providecommand \citenamefont [1]{#1}%
\providecommand \href@noop [0]{\@secondoftwo}%
\providecommand \href [0]{\begingroup \@sanitize@url \@href}%
\providecommand \@href[1]{\@@startlink{#1}\@@href}%
\providecommand \@@href[1]{\endgroup#1\@@endlink}%
\providecommand \@sanitize@url [0]{\catcode `\\12\catcode `\$12\catcode `\&12\catcode `\#12\catcode `\^12\catcode `\_12\catcode `\%12\relax}%
\providecommand \@@startlink[1]{}%
\providecommand \@@endlink[0]{}%
\providecommand \url  [0]{\begingroup\@sanitize@url \@url }%
\providecommand \@url [1]{\endgroup\@href {#1}{\urlprefix }}%
\providecommand \urlprefix  [0]{URL }%
\providecommand \Eprint [0]{\href }%
\providecommand \doibase [0]{https://doi.org/}%
\providecommand \selectlanguage [0]{\@gobble}%
\providecommand \bibinfo  [0]{\@secondoftwo}%
\providecommand \bibfield  [0]{\@secondoftwo}%
\providecommand \translation [1]{[#1]}%
\providecommand \BibitemOpen [0]{}%
\providecommand \bibitemStop [0]{}%
\providecommand \bibitemNoStop [0]{.\EOS\space}%
\providecommand \EOS [0]{\spacefactor3000\relax}%
\providecommand \BibitemShut  [1]{\csname bibitem#1\endcsname}%
\let\auto@bib@innerbib\@empty
\bibitem [{\citenamefont {Steinigeweg}\ \emph {et~al.}(2013)\citenamefont {Steinigeweg}, \citenamefont {Herbrych},\ and\ \citenamefont {Prelov{\v{s}}ek}}]{steinigeweg2013eigenstate}%
  \BibitemOpen
  \bibfield  {author} {\bibinfo {author} {\bibfnamefont {R.}~\bibnamefont {Steinigeweg}}, \bibinfo {author} {\bibfnamefont {J.}~\bibnamefont {Herbrych}},\ and\ \bibinfo {author} {\bibfnamefont {P.}~\bibnamefont {Prelov{\v{s}}ek}},\ }\bibfield  {title} {\bibinfo {title} {Eigenstate thermalization within isolated spin-chain systems},\ }\href@noop {} {\bibfield  {journal} {\bibinfo  {journal} {Physical Review E}\ }\textbf {\bibinfo {volume} {87}},\ \bibinfo {pages} {012118} (\bibinfo {year} {2013})}\BibitemShut {NoStop}%
\bibitem [{\citenamefont {Deutsch}(1991)}]{deutsch1991quantum}%
  \BibitemOpen
  \bibfield  {author} {\bibinfo {author} {\bibfnamefont {J.~M.}\ \bibnamefont {Deutsch}},\ }\bibfield  {title} {\bibinfo {title} {Quantum statistical mechanics in a closed system},\ }\href@noop {} {\bibfield  {journal} {\bibinfo  {journal} {Physical review a}\ }\textbf {\bibinfo {volume} {43}},\ \bibinfo {pages} {2046} (\bibinfo {year} {1991})}\BibitemShut {NoStop}%
\bibitem [{\citenamefont {Srednicki}(1994)}]{srednicki1994chaos}%
  \BibitemOpen
  \bibfield  {author} {\bibinfo {author} {\bibfnamefont {M.}~\bibnamefont {Srednicki}},\ }\bibfield  {title} {\bibinfo {title} {Chaos and quantum thermalization},\ }\href@noop {} {\bibfield  {journal} {\bibinfo  {journal} {Physical review e}\ }\textbf {\bibinfo {volume} {50}},\ \bibinfo {pages} {888} (\bibinfo {year} {1994})}\BibitemShut {NoStop}%
\bibitem [{\citenamefont {Calabrese}\ and\ \citenamefont {Cardy}(2004)}]{calabrese2004entanglement}%
  \BibitemOpen
  \bibfield  {author} {\bibinfo {author} {\bibfnamefont {P.}~\bibnamefont {Calabrese}}\ and\ \bibinfo {author} {\bibfnamefont {J.}~\bibnamefont {Cardy}},\ }\bibfield  {title} {\bibinfo {title} {Entanglement entropy and quantum field theory},\ }\href@noop {} {\bibfield  {journal} {\bibinfo  {journal} {Journal of statistical mechanics: theory and experiment}\ }\textbf {\bibinfo {volume} {2004}},\ \bibinfo {pages} {P06002} (\bibinfo {year} {2004})}\BibitemShut {NoStop}%
\bibitem [{\citenamefont {Eisert}\ \emph {et~al.}(2010)\citenamefont {Eisert}, \citenamefont {Cramer},\ and\ \citenamefont {Plenio}}]{eisert2010colloquium}%
  \BibitemOpen
  \bibfield  {author} {\bibinfo {author} {\bibfnamefont {J.}~\bibnamefont {Eisert}}, \bibinfo {author} {\bibfnamefont {M.}~\bibnamefont {Cramer}},\ and\ \bibinfo {author} {\bibfnamefont {M.~B.}\ \bibnamefont {Plenio}},\ }\bibfield  {title} {\bibinfo {title} {Colloquium: {A}rea laws for the entanglement entropy},\ }\href@noop {} {\bibfield  {journal} {\bibinfo  {journal} {Reviews of modern physics}\ }\textbf {\bibinfo {volume} {82}},\ \bibinfo {pages} {277} (\bibinfo {year} {2010})}\BibitemShut {NoStop}%
\bibitem [{\citenamefont {Calabrese}\ and\ \citenamefont {Cardy}(2005)}]{calabrese2005evolution}%
  \BibitemOpen
  \bibfield  {author} {\bibinfo {author} {\bibfnamefont {P.}~\bibnamefont {Calabrese}}\ and\ \bibinfo {author} {\bibfnamefont {J.}~\bibnamefont {Cardy}},\ }\bibfield  {title} {\bibinfo {title} {Evolution of entanglement entropy in one-dimensional systems},\ }\href@noop {} {\bibfield  {journal} {\bibinfo  {journal} {Journal of Statistical Mechanics: Theory and Experiment}\ }\textbf {\bibinfo {volume} {2005}},\ \bibinfo {pages} {P04010} (\bibinfo {year} {2005})}\BibitemShut {NoStop}%
\bibitem [{\citenamefont {Prosen}\ and\ \citenamefont {{\v{Z}}nidari{\v{c}}}(2009)}]{prosen2009matrix}%
  \BibitemOpen
  \bibfield  {author} {\bibinfo {author} {\bibfnamefont {T.}~\bibnamefont {Prosen}}\ and\ \bibinfo {author} {\bibfnamefont {M.}~\bibnamefont {{\v{Z}}nidari{\v{c}}}},\ }\bibfield  {title} {\bibinfo {title} {Matrix product simulations of non-equilibrium steady states of quantum spin chains},\ }\href@noop {} {\bibfield  {journal} {\bibinfo  {journal} {Journal of Statistical Mechanics: Theory and Experiment}\ }\textbf {\bibinfo {volume} {2009}},\ \bibinfo {pages} {P02035} (\bibinfo {year} {2009})}\BibitemShut {NoStop}%
\bibitem [{\citenamefont {Prosen}(2008)}]{prosen2008third}%
  \BibitemOpen
  \bibfield  {author} {\bibinfo {author} {\bibfnamefont {T.}~\bibnamefont {Prosen}},\ }\bibfield  {title} {\bibinfo {title} {Third quantization: a general method to solve master equations for quadratic open {F}ermi systems},\ }\href@noop {} {\bibfield  {journal} {\bibinfo  {journal} {New Journal of Physics}\ }\textbf {\bibinfo {volume} {10}},\ \bibinfo {pages} {043026} (\bibinfo {year} {2008})}\BibitemShut {NoStop}%
\bibitem [{\citenamefont {Prosen}(2011)}]{prosen2011open}%
  \BibitemOpen
  \bibfield  {author} {\bibinfo {author} {\bibfnamefont {T.}~\bibnamefont {Prosen}},\ }\bibfield  {title} {\bibinfo {title} {Open {X}{X}{Z} spin chain: {N}onequilibrium steady state and a strict bound on ballistic transport},\ }\href@noop {} {\bibfield  {journal} {\bibinfo  {journal} {Physical review letters}\ }\textbf {\bibinfo {volume} {106}},\ \bibinfo {pages} {217206} (\bibinfo {year} {2011})}\BibitemShut {NoStop}%
\bibitem [{\citenamefont {De~Nardis}\ \emph {et~al.}(2018)\citenamefont {De~Nardis}, \citenamefont {Bernard},\ and\ \citenamefont {Doyon}}]{de2018hydrodynamic}%
  \BibitemOpen
  \bibfield  {author} {\bibinfo {author} {\bibfnamefont {J.}~\bibnamefont {De~Nardis}}, \bibinfo {author} {\bibfnamefont {D.}~\bibnamefont {Bernard}},\ and\ \bibinfo {author} {\bibfnamefont {B.}~\bibnamefont {Doyon}},\ }\bibfield  {title} {\bibinfo {title} {Hydrodynamic diffusion in integrable systems},\ }\href@noop {} {\bibfield  {journal} {\bibinfo  {journal} {Physical review letters}\ }\textbf {\bibinfo {volume} {121}},\ \bibinfo {pages} {160603} (\bibinfo {year} {2018})}\BibitemShut {NoStop}%
\bibitem [{\citenamefont {{\v{Z}}nidari{\v{c}}}\ \emph {et~al.}(2016)\citenamefont {{\v{Z}}nidari{\v{c}}}, \citenamefont {Scardicchio},\ and\ \citenamefont {Varma}}]{vznidarivc2016diffusive}%
  \BibitemOpen
  \bibfield  {author} {\bibinfo {author} {\bibfnamefont {M.}~\bibnamefont {{\v{Z}}nidari{\v{c}}}}, \bibinfo {author} {\bibfnamefont {A.}~\bibnamefont {Scardicchio}},\ and\ \bibinfo {author} {\bibfnamefont {V.~K.}\ \bibnamefont {Varma}},\ }\bibfield  {title} {\bibinfo {title} {Diffusive and subdiffusive spin transport in the ergodic phase of a many-body localizable system},\ }\href@noop {} {\bibfield  {journal} {\bibinfo  {journal} {Physical review letters}\ }\textbf {\bibinfo {volume} {117}},\ \bibinfo {pages} {040601} (\bibinfo {year} {2016})}\BibitemShut {NoStop}%
\bibitem [{\citenamefont {Sologubenko}\ \emph {et~al.}(2007)\citenamefont {Sologubenko}, \citenamefont {Lorenz}, \citenamefont {Ott},\ and\ \citenamefont {Freimuth}}]{sologubenko2007thermal}%
  \BibitemOpen
  \bibfield  {author} {\bibinfo {author} {\bibfnamefont {A.}~\bibnamefont {Sologubenko}}, \bibinfo {author} {\bibfnamefont {T.}~\bibnamefont {Lorenz}}, \bibinfo {author} {\bibfnamefont {H.~R.}\ \bibnamefont {Ott}},\ and\ \bibinfo {author} {\bibfnamefont {A.}~\bibnamefont {Freimuth}},\ }\bibfield  {title} {\bibinfo {title} {Thermal conductivity via magnetic excitations in spin-chain materials},\ }\href@noop {} {\bibfield  {journal} {\bibinfo  {journal} {Journal of Low Temperature Physics}\ }\textbf {\bibinfo {volume} {147}},\ \bibinfo {pages} {387} (\bibinfo {year} {2007})}\BibitemShut {NoStop}%
\bibitem [{\citenamefont {Pan}\ \emph {et~al.}(2022)\citenamefont {Pan}, \citenamefont {Xu}, \citenamefont {Ni}, \citenamefont {Zhou}, \citenamefont {Hong}, \citenamefont {Qiu},\ and\ \citenamefont {Li}}]{pan2022unambiguous}%
  \BibitemOpen
  \bibfield  {author} {\bibinfo {author} {\bibfnamefont {B.}~\bibnamefont {Pan}}, \bibinfo {author} {\bibfnamefont {Y.}~\bibnamefont {Xu}}, \bibinfo {author} {\bibfnamefont {J.}~\bibnamefont {Ni}}, \bibinfo {author} {\bibfnamefont {S.}~\bibnamefont {Zhou}}, \bibinfo {author} {\bibfnamefont {X.}~\bibnamefont {Hong}}, \bibinfo {author} {\bibfnamefont {X.}~\bibnamefont {Qiu}},\ and\ \bibinfo {author} {\bibfnamefont {S.}~\bibnamefont {Li}},\ }\bibfield  {title} {\bibinfo {title} {Unambiguous experimental verification of linear-in-temperature spinon thermal conductivity in an antiferromagnetic {H}eisenberg chain},\ }\href@noop {} {\bibfield  {journal} {\bibinfo  {journal} {Physical Review Letters}\ }\textbf {\bibinfo {volume} {129}},\ \bibinfo {pages} {167201} (\bibinfo {year} {2022})}\BibitemShut {NoStop}%
\bibitem [{\citenamefont {Hlubek}\ \emph {et~al.}(2010)\citenamefont {Hlubek}, \citenamefont {Ribeiro}, \citenamefont {Saint-Martin}, \citenamefont {Revcolevschi}, \citenamefont {Roth}, \citenamefont {Behr}, \citenamefont {B{\"u}chner},\ and\ \citenamefont {Hess}}]{hlubek2010ballistic}%
  \BibitemOpen
  \bibfield  {author} {\bibinfo {author} {\bibfnamefont {N.}~\bibnamefont {Hlubek}}, \bibinfo {author} {\bibfnamefont {P.}~\bibnamefont {Ribeiro}}, \bibinfo {author} {\bibfnamefont {R.}~\bibnamefont {Saint-Martin}}, \bibinfo {author} {\bibfnamefont {A.}~\bibnamefont {Revcolevschi}}, \bibinfo {author} {\bibfnamefont {G.}~\bibnamefont {Roth}}, \bibinfo {author} {\bibfnamefont {G.}~\bibnamefont {Behr}}, \bibinfo {author} {\bibfnamefont {B.}~\bibnamefont {B{\"u}chner}},\ and\ \bibinfo {author} {\bibfnamefont {C.}~\bibnamefont {Hess}},\ }\bibfield  {title} {\bibinfo {title} {Ballistic heat transport of quantum spin excitations as seen in $\text{SrCuO}_2$},\ }\href@noop {} {\bibfield  {journal} {\bibinfo  {journal} {Physical Review B}\ }\textbf {\bibinfo {volume} {81}},\ \bibinfo {pages} {020405} (\bibinfo {year} {2010})}\BibitemShut {NoStop}%
\bibitem [{\citenamefont {Karrasch}\ \emph {et~al.}(2015)\citenamefont {Karrasch}, \citenamefont {Kennes},\ and\ \citenamefont {Heidrich-Meisner}}]{karrasch2015spin}%
  \BibitemOpen
  \bibfield  {author} {\bibinfo {author} {\bibfnamefont {C.}~\bibnamefont {Karrasch}}, \bibinfo {author} {\bibfnamefont {D.}~\bibnamefont {Kennes}},\ and\ \bibinfo {author} {\bibfnamefont {F.}~\bibnamefont {Heidrich-Meisner}},\ }\bibfield  {title} {\bibinfo {title} {Spin and thermal conductivity of quantum spin chains and ladders},\ }\href@noop {} {\bibfield  {journal} {\bibinfo  {journal} {Physical Review B}\ }\textbf {\bibinfo {volume} {91}},\ \bibinfo {pages} {115130} (\bibinfo {year} {2015})}\BibitemShut {NoStop}%
\bibitem [{\citenamefont {Zhang}\ \emph {et~al.}(2008)\citenamefont {Zhang}, \citenamefont {Wang},\ and\ \citenamefont {Li}}]{zhang2008ballistic}%
  \BibitemOpen
  \bibfield  {author} {\bibinfo {author} {\bibfnamefont {L.}~\bibnamefont {Zhang}}, \bibinfo {author} {\bibfnamefont {J.-S.}\ \bibnamefont {Wang}},\ and\ \bibinfo {author} {\bibfnamefont {B.}~\bibnamefont {Li}},\ }\bibfield  {title} {\bibinfo {title} {Ballistic magnetothermal transport in a {H}eisenberg spin chain at low temperatures},\ }\href@noop {} {\bibfield  {journal} {\bibinfo  {journal} {Physical Review B}\ }\textbf {\bibinfo {volume} {78}},\ \bibinfo {pages} {144416} (\bibinfo {year} {2008})}\BibitemShut {NoStop}%
\bibitem [{\citenamefont {Laflorencie}\ \emph {et~al.}(2004)\citenamefont {Laflorencie}, \citenamefont {Rieger}, \citenamefont {Sandvik},\ and\ \citenamefont {Henelius}}]{laflorencie2004crossover}%
  \BibitemOpen
  \bibfield  {author} {\bibinfo {author} {\bibfnamefont {N.}~\bibnamefont {Laflorencie}}, \bibinfo {author} {\bibfnamefont {H.}~\bibnamefont {Rieger}}, \bibinfo {author} {\bibfnamefont {A.~W.}\ \bibnamefont {Sandvik}},\ and\ \bibinfo {author} {\bibfnamefont {P.}~\bibnamefont {Henelius}},\ }\bibfield  {title} {\bibinfo {title} {Crossover effects in the random-exchange spin-$1/2$ antiferromagnetic chain},\ }\href@noop {} {\bibfield  {journal} {\bibinfo  {journal} {Physical Review B}\ }\textbf {\bibinfo {volume} {70}},\ \bibinfo {pages} {054430} (\bibinfo {year} {2004})}\BibitemShut {NoStop}%
\bibitem [{\citenamefont {Heidrich-Meisner}\ \emph {et~al.}(2005)\citenamefont {Heidrich-Meisner}, \citenamefont {Honecker},\ and\ \citenamefont {Brenig}}]{heidrich2005thermal}%
  \BibitemOpen
  \bibfield  {author} {\bibinfo {author} {\bibfnamefont {F.}~\bibnamefont {Heidrich-Meisner}}, \bibinfo {author} {\bibfnamefont {A.}~\bibnamefont {Honecker}},\ and\ \bibinfo {author} {\bibfnamefont {W.}~\bibnamefont {Brenig}},\ }\bibfield  {title} {\bibinfo {title} {Thermal transport of the {X}{X}{Z} chain in a magnetic field},\ }\href@noop {} {\bibfield  {journal} {\bibinfo  {journal} {Physical Review B}\ }\textbf {\bibinfo {volume} {71}},\ \bibinfo {pages} {184415} (\bibinfo {year} {2005})}\BibitemShut {NoStop}%
\bibitem [{\citenamefont {Mendoza-Arenas}\ \emph {et~al.}(2013)\citenamefont {Mendoza-Arenas}, \citenamefont {Al-Assam}, \citenamefont {Clark},\ and\ \citenamefont {Jaksch}}]{mendoza2013heat}%
  \BibitemOpen
  \bibfield  {author} {\bibinfo {author} {\bibfnamefont {J.}~\bibnamefont {Mendoza-Arenas}}, \bibinfo {author} {\bibfnamefont {S.}~\bibnamefont {Al-Assam}}, \bibinfo {author} {\bibfnamefont {S.}~\bibnamefont {Clark}},\ and\ \bibinfo {author} {\bibfnamefont {D.}~\bibnamefont {Jaksch}},\ }\bibfield  {title} {\bibinfo {title} {Heat transport in the {X}{X}{Z} spin chain: from ballistic to diffusive regimes and dephasing enhancement},\ }\href@noop {} {\bibfield  {journal} {\bibinfo  {journal} {Journal of Statistical Mechanics: Theory and Experiment}\ }\textbf {\bibinfo {volume} {2013}},\ \bibinfo {pages} {P07007} (\bibinfo {year} {2013})}\BibitemShut {NoStop}%
\bibitem [{\citenamefont {Lange}\ \emph {et~al.}(2019)\citenamefont {Lange}, \citenamefont {Ejima},\ and\ \citenamefont {Fehske}}]{lange2019driving}%
  \BibitemOpen
  \bibfield  {author} {\bibinfo {author} {\bibfnamefont {F.}~\bibnamefont {Lange}}, \bibinfo {author} {\bibfnamefont {S.}~\bibnamefont {Ejima}},\ and\ \bibinfo {author} {\bibfnamefont {H.}~\bibnamefont {Fehske}},\ }\bibfield  {title} {\bibinfo {title} {Driving {X}{X}{Z} spin chains: {M}agnetic-field and boundary effects},\ }\href@noop {} {\bibfield  {journal} {\bibinfo  {journal} {Europhysics Letters}\ }\textbf {\bibinfo {volume} {125}},\ \bibinfo {pages} {17001} (\bibinfo {year} {2019})}\BibitemShut {NoStop}%
\bibitem [{\citenamefont {Lenar{\v{c}}i{\v{c}}}\ and\ \citenamefont {Prosen}(2015)}]{lenarvcivc2015exact}%
  \BibitemOpen
  \bibfield  {author} {\bibinfo {author} {\bibfnamefont {Z.}~\bibnamefont {Lenar{\v{c}}i{\v{c}}}}\ and\ \bibinfo {author} {\bibfnamefont {T.}~\bibnamefont {Prosen}},\ }\bibfield  {title} {\bibinfo {title} {Exact asymptotics of the current in boundary-driven dissipative quantum chains in large external fields},\ }\href@noop {} {\bibfield  {journal} {\bibinfo  {journal} {Physical Review E}\ }\textbf {\bibinfo {volume} {91}},\ \bibinfo {pages} {030103} (\bibinfo {year} {2015})}\BibitemShut {NoStop}%
\bibitem [{\citenamefont {Schulz}\ \emph {et~al.}(2018)\citenamefont {Schulz}, \citenamefont {Taylor}, \citenamefont {Hooley},\ and\ \citenamefont {Scardicchio}}]{schulz2018energy}%
  \BibitemOpen
  \bibfield  {author} {\bibinfo {author} {\bibfnamefont {M.}~\bibnamefont {Schulz}}, \bibinfo {author} {\bibfnamefont {S.~R.}\ \bibnamefont {Taylor}}, \bibinfo {author} {\bibfnamefont {C.~A.}\ \bibnamefont {Hooley}},\ and\ \bibinfo {author} {\bibfnamefont {A.}~\bibnamefont {Scardicchio}},\ }\bibfield  {title} {\bibinfo {title} {Energy transport in a disordered spin chain with broken $u(1)$ symmetry: {D}iffusion, subdiffusion, and many-body localization},\ }\href@noop {} {\bibfield  {journal} {\bibinfo  {journal} {Physical Review B}\ }\textbf {\bibinfo {volume} {98}},\ \bibinfo {pages} {180201} (\bibinfo {year} {2018})}\BibitemShut {NoStop}%
\bibitem [{\citenamefont {{\v{Z}}nidari{\v{c}}}\ and\ \citenamefont {Horvat}(2013)}]{vznidarivc2013transport}%
  \BibitemOpen
  \bibfield  {author} {\bibinfo {author} {\bibfnamefont {M.}~\bibnamefont {{\v{Z}}nidari{\v{c}}}}\ and\ \bibinfo {author} {\bibfnamefont {M.}~\bibnamefont {Horvat}},\ }\bibfield  {title} {\bibinfo {title} {Transport in a disordered tight-binding chain with dephasing},\ }\href@noop {} {\bibfield  {journal} {\bibinfo  {journal} {The European Physical Journal B}\ }\textbf {\bibinfo {volume} {86}},\ \bibinfo {pages} {1} (\bibinfo {year} {2013})}\BibitemShut {NoStop}%
\bibitem [{\citenamefont {Landi}\ \emph {et~al.}(2014)\citenamefont {Landi}, \citenamefont {Novais}, \citenamefont {De~Oliveira},\ and\ \citenamefont {Karevski}}]{landi2014flux}%
  \BibitemOpen
  \bibfield  {author} {\bibinfo {author} {\bibfnamefont {G.~T.}\ \bibnamefont {Landi}}, \bibinfo {author} {\bibfnamefont {E.}~\bibnamefont {Novais}}, \bibinfo {author} {\bibfnamefont {M.~J.}\ \bibnamefont {De~Oliveira}},\ and\ \bibinfo {author} {\bibfnamefont {D.}~\bibnamefont {Karevski}},\ }\bibfield  {title} {\bibinfo {title} {Flux rectification in the quantum {X}{X}{Z} chain},\ }\href@noop {} {\bibfield  {journal} {\bibinfo  {journal} {Physical Review E}\ }\textbf {\bibinfo {volume} {90}},\ \bibinfo {pages} {042142} (\bibinfo {year} {2014})}\BibitemShut {NoStop}%
\bibitem [{\citenamefont {Kl{\"u}mper}\ and\ \citenamefont {Sakai}(2002)}]{klumper2002thermal}%
  \BibitemOpen
  \bibfield  {author} {\bibinfo {author} {\bibfnamefont {A.}~\bibnamefont {Kl{\"u}mper}}\ and\ \bibinfo {author} {\bibfnamefont {K.}~\bibnamefont {Sakai}},\ }\bibfield  {title} {\bibinfo {title} {The thermal conductivity of the spin-$1/2$ {X}{X}{Z} chain at arbitrary temperature},\ }\href@noop {} {\bibfield  {journal} {\bibinfo  {journal} {Journal of Physics A: Mathematical and General}\ }\textbf {\bibinfo {volume} {35}},\ \bibinfo {pages} {2173} (\bibinfo {year} {2002})}\BibitemShut {NoStop}%
\bibitem [{\citenamefont {Langer}\ \emph {et~al.}(2009)\citenamefont {Langer}, \citenamefont {Heidrich-Meisner}, \citenamefont {Gemmer}, \citenamefont {McCulloch},\ and\ \citenamefont {Schollw{\"o}ck}}]{langer2009real}%
  \BibitemOpen
  \bibfield  {author} {\bibinfo {author} {\bibfnamefont {S.}~\bibnamefont {Langer}}, \bibinfo {author} {\bibfnamefont {F.}~\bibnamefont {Heidrich-Meisner}}, \bibinfo {author} {\bibfnamefont {J.}~\bibnamefont {Gemmer}}, \bibinfo {author} {\bibfnamefont {I.}~\bibnamefont {McCulloch}},\ and\ \bibinfo {author} {\bibfnamefont {U.}~\bibnamefont {Schollw{\"o}ck}},\ }\bibfield  {title} {\bibinfo {title} {Real-time study of diffusive and ballistic transport in spin-$1/2$ chains using the adaptive time-dependent density matrix renormalization group method},\ }\href@noop {} {\bibfield  {journal} {\bibinfo  {journal} {Physical Review B}\ }\textbf {\bibinfo {volume} {79}},\ \bibinfo {pages} {214409} (\bibinfo {year} {2009})}\BibitemShut {NoStop}%
\bibitem [{\citenamefont {Biella}\ \emph {et~al.}(2019)\citenamefont {Biella}, \citenamefont {Collura}, \citenamefont {Rossini}, \citenamefont {De~Luca},\ and\ \citenamefont {Mazza}}]{biella2019ballistic}%
  \BibitemOpen
  \bibfield  {author} {\bibinfo {author} {\bibfnamefont {A.}~\bibnamefont {Biella}}, \bibinfo {author} {\bibfnamefont {M.}~\bibnamefont {Collura}}, \bibinfo {author} {\bibfnamefont {D.}~\bibnamefont {Rossini}}, \bibinfo {author} {\bibfnamefont {A.}~\bibnamefont {De~Luca}},\ and\ \bibinfo {author} {\bibfnamefont {L.}~\bibnamefont {Mazza}},\ }\bibfield  {title} {\bibinfo {title} {Ballistic transport and boundary resistances in inhomogeneous quantum spin chains},\ }\href@noop {} {\bibfield  {journal} {\bibinfo  {journal} {Nature communications}\ }\textbf {\bibinfo {volume} {10}},\ \bibinfo {pages} {4820} (\bibinfo {year} {2019})}\BibitemShut {NoStop}%
\bibitem [{\citenamefont {Benatti}\ \emph {et~al.}(2021)\citenamefont {Benatti}, \citenamefont {Floreanini},\ and\ \citenamefont {Memarzadeh}}]{benatti2021exact}%
  \BibitemOpen
  \bibfield  {author} {\bibinfo {author} {\bibfnamefont {F.}~\bibnamefont {Benatti}}, \bibinfo {author} {\bibfnamefont {R.}~\bibnamefont {Floreanini}},\ and\ \bibinfo {author} {\bibfnamefont {L.}~\bibnamefont {Memarzadeh}},\ }\bibfield  {title} {\bibinfo {title} {Exact {S}teady {S}tate of the {O}pen {X}{X}-{S}pin {C}hain: {E}ntanglement and {T}ransport {P}roperties},\ }\href@noop {} {\bibfield  {journal} {\bibinfo  {journal} {PRX Quantum}\ }\textbf {\bibinfo {volume} {2}},\ \bibinfo {pages} {030344} (\bibinfo {year} {2021})}\BibitemShut {NoStop}%
\bibitem [{\citenamefont {Benatti}\ \emph {et~al.}(2022)\citenamefont {Benatti}, \citenamefont {Floreanini},\ and\ \citenamefont {Memarzadeh}}]{benatti2022stationary}%
  \BibitemOpen
  \bibfield  {author} {\bibinfo {author} {\bibfnamefont {F.}~\bibnamefont {Benatti}}, \bibinfo {author} {\bibfnamefont {R.}~\bibnamefont {Floreanini}},\ and\ \bibinfo {author} {\bibfnamefont {L.}~\bibnamefont {Memarzadeh}},\ }\bibfield  {title} {\bibinfo {title} {Stationary states of open {X}{X}-spin chains},\ }\href@noop {} {\bibfield  {journal} {\bibinfo  {journal} {Physical Review A}\ }\textbf {\bibinfo {volume} {106}},\ \bibinfo {pages} {062218} (\bibinfo {year} {2022})}\BibitemShut {NoStop}%
\bibitem [{\citenamefont {Christandl}\ \emph {et~al.}(2004)\citenamefont {Christandl}, \citenamefont {Datta}, \citenamefont {Ekert},\ and\ \citenamefont {Landahl}}]{christandl2004perfect}%
  \BibitemOpen
  \bibfield  {author} {\bibinfo {author} {\bibfnamefont {M.}~\bibnamefont {Christandl}}, \bibinfo {author} {\bibfnamefont {N.}~\bibnamefont {Datta}}, \bibinfo {author} {\bibfnamefont {A.}~\bibnamefont {Ekert}},\ and\ \bibinfo {author} {\bibfnamefont {A.~J.}\ \bibnamefont {Landahl}},\ }\bibfield  {title} {\bibinfo {title} {Perfect state transfer in quantum spin networks},\ }\href@noop {} {\bibfield  {journal} {\bibinfo  {journal} {Physical review letters}\ }\textbf {\bibinfo {volume} {92}},\ \bibinfo {pages} {187902} (\bibinfo {year} {2004})}\BibitemShut {NoStop}%
\bibitem [{\citenamefont {Vinet}\ and\ \citenamefont {Zhedanov}(2012{\natexlab{a}})}]{howToConstruct}%
  \BibitemOpen
  \bibfield  {author} {\bibinfo {author} {\bibfnamefont {L.}~\bibnamefont {Vinet}}\ and\ \bibinfo {author} {\bibfnamefont {A.}~\bibnamefont {Zhedanov}},\ }\bibfield  {title} {\bibinfo {title} {How to construct spin chains with perfect state transfer},\ }\bibfield  {journal} {\bibinfo  {journal} {Physical Review A}\ }\textbf {\bibinfo {volume} {85}},\ \href {https://doi.org/10.1103/physreva.85.012323} {10.1103/physreva.85.012323} (\bibinfo {year} {2012}{\natexlab{a}})\BibitemShut {NoStop}%
\bibitem [{\citenamefont {Cramp{\'e}}\ \emph {et~al.}(2021)\citenamefont {Cramp{\'e}}, \citenamefont {Nepomechie},\ and\ \citenamefont {Vinet}}]{crampe2021entanglement}%
  \BibitemOpen
  \bibfield  {author} {\bibinfo {author} {\bibfnamefont {N.}~\bibnamefont {Cramp{\'e}}}, \bibinfo {author} {\bibfnamefont {R.~I.}\ \bibnamefont {Nepomechie}},\ and\ \bibinfo {author} {\bibfnamefont {L.}~\bibnamefont {Vinet}},\ }\bibfield  {title} {\bibinfo {title} {Entanglement in fermionic chains and bispectrality},\ }\href@noop {} {\bibfield  {journal} {\bibinfo  {journal} {Reviews in Mathematical Physics}\ }\textbf {\bibinfo {volume} {33}},\ \bibinfo {pages} {2140001} (\bibinfo {year} {2021})}\BibitemShut {NoStop}%
\bibitem [{\citenamefont {Koekoek}\ \emph {et~al.}(2010)\citenamefont {Koekoek}, \citenamefont {Lesky}, \citenamefont {Swarttouw}, \citenamefont {Koekoek}, \citenamefont {Lesky},\ and\ \citenamefont {Swarttouw}}]{koekoek}%
  \BibitemOpen
  \bibfield  {author} {\bibinfo {author} {\bibfnamefont {R.}~\bibnamefont {Koekoek}}, \bibinfo {author} {\bibfnamefont {P.~A.}\ \bibnamefont {Lesky}}, \bibinfo {author} {\bibfnamefont {R.~F.}\ \bibnamefont {Swarttouw}}, \bibinfo {author} {\bibfnamefont {R.}~\bibnamefont {Koekoek}}, \bibinfo {author} {\bibfnamefont {P.~A.}\ \bibnamefont {Lesky}},\ and\ \bibinfo {author} {\bibfnamefont {R.~F.}\ \bibnamefont {Swarttouw}},\ }\href@noop {} {\emph {\bibinfo {title} {Hypergeometric orthogonal polynomials}}}\ (\bibinfo  {publisher} {Springer},\ \bibinfo {year} {2010})\BibitemShut {NoStop}%
\bibitem [{\citenamefont {Kay}(2010)}]{kay2010perfect}%
  \BibitemOpen
  \bibfield  {author} {\bibinfo {author} {\bibfnamefont {A.}~\bibnamefont {Kay}},\ }\bibfield  {title} {\bibinfo {title} {Perfect, efficient, state transfer and its application as a constructive tool},\ }\href@noop {} {\bibfield  {journal} {\bibinfo  {journal} {International Journal of Quantum Information}\ }\textbf {\bibinfo {volume} {8}},\ \bibinfo {pages} {641} (\bibinfo {year} {2010})}\BibitemShut {NoStop}%
\bibitem [{\citenamefont {Albanese}\ \emph {et~al.}(2004)\citenamefont {Albanese}, \citenamefont {Christandl}, \citenamefont {Datta},\ and\ \citenamefont {Ekert}}]{albanese2004mirror}%
  \BibitemOpen
  \bibfield  {author} {\bibinfo {author} {\bibfnamefont {C.}~\bibnamefont {Albanese}}, \bibinfo {author} {\bibfnamefont {M.}~\bibnamefont {Christandl}}, \bibinfo {author} {\bibfnamefont {N.}~\bibnamefont {Datta}},\ and\ \bibinfo {author} {\bibfnamefont {A.}~\bibnamefont {Ekert}},\ }\bibfield  {title} {\bibinfo {title} {Mirror inversion of quantum states in linear registers},\ }\href@noop {} {\bibfield  {journal} {\bibinfo  {journal} {Physical review letters}\ }\textbf {\bibinfo {volume} {93}},\ \bibinfo {pages} {230502} (\bibinfo {year} {2004})}\BibitemShut {NoStop}%
\bibitem [{\citenamefont {Liu}\ \emph {et~al.}(2014)\citenamefont {Liu}, \citenamefont {H{\"a}nggi}, \citenamefont {Li}, \citenamefont {Ren},\ and\ \citenamefont {Li}}]{liu2014anomalous}%
  \BibitemOpen
  \bibfield  {author} {\bibinfo {author} {\bibfnamefont {S.}~\bibnamefont {Liu}}, \bibinfo {author} {\bibfnamefont {P.}~\bibnamefont {H{\"a}nggi}}, \bibinfo {author} {\bibfnamefont {N.}~\bibnamefont {Li}}, \bibinfo {author} {\bibfnamefont {J.}~\bibnamefont {Ren}},\ and\ \bibinfo {author} {\bibfnamefont {B.}~\bibnamefont {Li}},\ }\bibfield  {title} {\bibinfo {title} {Anomalous heat diffusion},\ }\href@noop {} {\bibfield  {journal} {\bibinfo  {journal} {Physical review letters}\ }\textbf {\bibinfo {volume} {112}},\ \bibinfo {pages} {040601} (\bibinfo {year} {2014})}\BibitemShut {NoStop}%
\bibitem [{\citenamefont {Benatti}\ and\ \citenamefont {Floreanini}(2005)}]{BENATTI_2005}%
  \BibitemOpen
  \bibfield  {author} {\bibinfo {author} {\bibfnamefont {F.}~\bibnamefont {Benatti}}\ and\ \bibinfo {author} {\bibfnamefont {R.}~\bibnamefont {Floreanini}},\ }\bibfield  {title} {\bibinfo {title} {Open {Q}unatum {D}ynamics: {C}omplete {P}ositivity and {E}ntanglement},\ }\href {https://doi.org/10.1142/s0217979205032097} {\bibfield  {journal} {\bibinfo  {journal} {International Journal of Modern Physics B}\ }\textbf {\bibinfo {volume} {19}},\ \bibinfo {pages} {3063–3139} (\bibinfo {year} {2005})}\BibitemShut {NoStop}%
\bibitem [{\citenamefont {Rivas}\ and\ \citenamefont {Huelga}(2012)}]{Rivas_2012}%
  \BibitemOpen
  \bibfield  {author} {\bibinfo {author} {\bibfnamefont {{\'A}.}~\bibnamefont {Rivas}}\ and\ \bibinfo {author} {\bibfnamefont {S.~F.}\ \bibnamefont {Huelga}},\ }\href {https://doi.org/10.1007/978-3-642-23354-8} {\emph {\bibinfo {title} {Open Quantum Systems: An Introduction}}}\ (\bibinfo  {publisher} {Springer Berlin Heidelberg},\ \bibinfo {year} {2012})\BibitemShut {NoStop}%
\bibitem [{\citenamefont {Alicki}\ and\ \citenamefont {Lendi}(2007)}]{Alicki2007}%
  \BibitemOpen
  \bibfield  {author} {\bibinfo {author} {\bibfnamefont {R.}~\bibnamefont {Alicki}}\ and\ \bibinfo {author} {\bibfnamefont {K.}~\bibnamefont {Lendi}},\ }\href {https://doi.org/10.1007/3-540-70861-8_3} {\emph {\bibinfo {title} {Quantum Dynamical Semigroups and Applications}}}\ (\bibinfo  {publisher} {Springer Berlin Heidelberg},\ \bibinfo {year} {2007})\BibitemShut {NoStop}%
\bibitem [{\citenamefont {Hess}(2007)}]{hess2007heat}%
  \BibitemOpen
  \bibfield  {author} {\bibinfo {author} {\bibfnamefont {C.}~\bibnamefont {Hess}},\ }\bibfield  {title} {\bibinfo {title} {Heat conduction in low-dimensional quantum magnets},\ }\href@noop {} {\bibfield  {journal} {\bibinfo  {journal} {The European Physical Journal Special Topics}\ }\textbf {\bibinfo {volume} {151}},\ \bibinfo {pages} {73} (\bibinfo {year} {2007})}\BibitemShut {NoStop}%
\bibitem [{\citenamefont {Sologubenko}\ \emph {et~al.}(2001)\citenamefont {Sologubenko}, \citenamefont {Gianno}, \citenamefont {Ott}, \citenamefont {Vietkine},\ and\ \citenamefont {Revcolevschi}}]{sologubenko2001heat}%
  \BibitemOpen
  \bibfield  {author} {\bibinfo {author} {\bibfnamefont {A.}~\bibnamefont {Sologubenko}}, \bibinfo {author} {\bibfnamefont {K.}~\bibnamefont {Gianno}}, \bibinfo {author} {\bibfnamefont {H.}~\bibnamefont {Ott}}, \bibinfo {author} {\bibfnamefont {A.}~\bibnamefont {Vietkine}},\ and\ \bibinfo {author} {\bibfnamefont {A.}~\bibnamefont {Revcolevschi}},\ }\bibfield  {title} {\bibinfo {title} {Heat transport by lattice and spin excitations in the spin-chain compounds $\text{SrCuO}_2$ and $\text{Sr}_2\text{CuO}_3$},\ }\href@noop {} {\bibfield  {journal} {\bibinfo  {journal} {Physical Review B}\ }\textbf {\bibinfo {volume} {64}},\ \bibinfo {pages} {054412} (\bibinfo {year} {2001})}\BibitemShut {NoStop}%
\bibitem [{\citenamefont {Vinet}\ and\ \citenamefont {Zhedanov}(2012{\natexlab{b}})}]{vinet2012construct}%
  \BibitemOpen
  \bibfield  {author} {\bibinfo {author} {\bibfnamefont {L.}~\bibnamefont {Vinet}}\ and\ \bibinfo {author} {\bibfnamefont {A.}~\bibnamefont {Zhedanov}},\ }\bibfield  {title} {\bibinfo {title} {How to construct spin chains with perfect state transfer},\ }\href@noop {} {\bibfield  {journal} {\bibinfo  {journal} {Physical Review A—Atomic, Molecular, and Optical Physics}\ }\textbf {\bibinfo {volume} {85}},\ \bibinfo {pages} {012323} (\bibinfo {year} {2012}{\natexlab{b}})}\BibitemShut {NoStop}%
\end{thebibliography}%

\end{document}